\documentclass[10pt,a4paper]{article}
\usepackage[utf8]{inputenc}
\usepackage{amsmath}
\usepackage{amsfonts}
\usepackage{amssymb}
\usepackage{caption}
\usepackage{graphicx}
\usepackage[Conny]{fncychap}
\usepackage[top=1in,bottom=1in,right=1in,left=1in,heightrounded]{geometry}
\title{Reinterpreting Landauer conductance, solving the quantum measurement problem, grand unification}
\author{Kanchan Meena\thanks{Gmail: 1996.kanchanmeena@gmail.com}, 
Souvik Ghosh\thanks{Gmail: souvikghosh2012@gmail.com} and P. Singha Deo\thanks{Gmail: singhadeoprosenjit@gmail.com}}
\date{
Chemical and biological physics department, Weizmann Institute of Science, 76100 Rehobot, Israel.\\
Department of Physics, National Sun Yat-sen University,No. 70, Lienhai Road, Kaohsiung 80424, Taiwan.\\
Department of physics of complex systems, SN Bose Centre, JD Block, Sec 3, Salt Lake, Kolkata, India 7000106.}
\begin{document}
\maketitle

\begin{abstract}

In a series of recent papers we have proved rigorously that time travel is
a reality and very much feasible by using quantum mechanical processes.
There are plenty of indirect experimental support until a direct experiment is conducted.
The process crucially depend on the reality of a local time as well as a local partial
density of states (LPDOS) that can become negative very easily in the quantum regime of mesoscopic systems.
Mesoscopic systems are small enough to allow us to experimentally access the intermediate regime between the classical and quantum worlds.
This LPDOS is in every sense a hidden variable in quantum mechanics that does not
show up in the axiomatic framework of quantum mechanics.
It can be inferred through physical clocks obeying quantum dynamics and can be rigorously justified from
the properties of the Hilbert space that is uniquely isomorphic to 
the complex plane.
Therefore one can naturally guess that LPDOS will have something important to say about quantum measurement
as well as the unification of classical and quantum laws.
We therefore undertake the exercise to
show that LPDOS can very much allow us to re-interpret the enormously successful phenomenological Landauer-Buttiker
formalism for mesoscopic systems and put it on firm theoretical ground as a bridge between classical
and quantum mechanics, thereby unifying them. Essentially the local time calculated quantum mechanically
can dilate exactly like the proper time
of relativity and be consistent with the coordinate time of relativity.
Also the measured conductance of mesoscopic
samples is a deterministic quantum measurement outcome from a linear superposition of states,
essentially because of LPDOS, which
solves the quantum measurement problem.
For this we analyze the three probe conductance formula in details and give our arguments for the general case.
\end{abstract}

\section{Introduction}

Mesoscopic regime is that in between the quantum and classical regimes due to the small size of the samples and
the low temperatures at which these systems are subjected to. Recent advances in fabrication techniques has allowed
us to make high mobility two dimensional electron gas (2DEG) on which lithographic techniques can be applied to fabricate
a small quantum system connected to leads that connect the quantum system to the classical world
\cite{dat} and that is an essential part of these
systems. Supposing these leads are labeled $\alpha$, $\beta$, $\gamma \;\;\cdots$, etc. One can make the systems
so small that at low temperatures the de Broglie wavelength of the electrons become comparable to the sample size and we
can look into quantum effects that are not subject to thermalization and ensemble averaging. Thus we get, what we call
a general mesoscopic system that help us probe deep into the interplay of classical and quantum laws. We
call it general because such a mesoscopic system
becomes a true ambassador to understand the intermediate regime that lies between the quantum and classical laws.
Starting from such a system, we can find variants of it to probe other problems in physics that may tend towards
the classical or quantum regimes.

Motivated by the developments in fabrication of such samples, there was a flurry of theoretical activity, that culminated
in the celebrated Landauer-Buttiker formalism to understand these small systems. The Landauer
conductance formula for example, is extremely successful in giving the conductance of such
small systems, that could even be a molecule connected to leads. However, conceptual problems
remain and we describe below a re-interpretation of the Landauer conductance, from the view point of our recent works \cite{kan}
and mentioned in the next paragraphs.

One cannot define a self adjoint time operator in quantum mechanics and the time that appears in Schrodinger equation
is just a parameter that is not directly related to what we may call measured time.
Landauer, Buttiker and others proposed the idea of a local state in such a mesoscopic system based on the idea of a physical
clock \cite{lar} that can give a local time. Since such a local time also means that there is a local state, it
contradicts the axioms of quantum mechanics.
The physical clock as envisaged by Landauer and Buttiker uses the idea that the spin of an electron can be
thought of as a classical magnetic dipole and if it is subjected to a classical force will precess like a classical dipole
in a magnetic field. The classical force can be obtained from a perturbation potential of the form $\vec \mu \cdot \vec B(\vec r)$ rather than
$\vec \mu \cdot \vec B(\hat r)$ which essentially means the dot product is not the inner product of quantum mechanics
but the usual dot product for a classical precession with an angle that changes with time.
That in turn means usual perturbation expansion in quantum mechanics is replaced by a direct Taylor expansion
of the asymptotic wave-function (or scattering matrix elements) followed by calculation of expectation value of
angular displacement with respect to asymptotic wave-function.
The magnitude of the magnetic moment $\vec \mu$ is
however taken to be quantized in units of $\frac{\hbar}{2}$.
The angular displacement expectation value divided by the classical Larmor frequency of a dipole of strength $\frac{\hbar}{2}$
then defines a time called Larmor precession time (LPT) for which a detailed derivation can be found in
\cite{kan}. The equation of motion for the electron, that is the Schrodinger equation,
is not used therefore, but only the analyticity of the spinor space is used.
There is no direct way to verify this object called LPT from the wave-function as will be explained, but if it is averaged over all
coordinates of the sample (that is averaged over $\textbf r$) and appropriate outgoing channels $\alpha$
then one arrives at an object called injectance of lead $\gamma$ which is completely in agreement with what one can calculate from the
wave-function. There remains the doubt whether the non-averaged object called LPT from which
one can also define a local partial density of states (LPDOS) $\rho_{lpd}$,
at all has any relevance to reality, specially since axiomatic framework of quantum mechanics denies its possibility.

Our works \cite{kan} remove all the above mentioned assumptions needed to derive LPDOS or LPT.
We show that the definition of this object $\rho_{lpd}$ via a physical clock is a natural consequence of the fact that an open system
is more general than a closed system. Meaning, it can only be defined if we have the leads $\gamma$, $\alpha$ etc, with their
open ends connected to classical electron reservoirs.
The classical reservoirs destroy the linear superposition of states in the leads while linear superposition
exists inside the sample (shaded region in Fig. 2) \cite{kan}.
Events in the reservoir are all classical events and quantum
event can only be defined in between two classical events in the reservoirs.
The closed system can be seen as a limiting case when the coupling to the leads tends to zero.
In that situation it can be shown that the object LPDOS defined in Eq. 56 follow directly from the topology of the complex plane
that is isomorphic to the Hilbert space (or the spinor space). The exact form of the term $\vec \mu \cdot \vec B(\hat r)$ for
which the inner product in Hilbert space replaces the notion of dot product in real space, does not matter.
That also naturally explains the local states with an LPDOS cannot be defined within the axiomatic framework
of quantum mechanics
because it is referring
to only those  electrons that are coming specifically from $\gamma$ and going specifically to $\alpha$. The wave-function
of an electron at a point $\textbf r$ can only be sensitive to either $\gamma$ or $\alpha$.
This over-specified object $\rho_{lpd}$ can be calculated for a mesoscopic system
from scattering matrix elements.
If we start from a closed system governed by Schrodinger equation, we cannot get an object that is $\gamma$ and
$\alpha$ dependent because in that case the idea of a physical clock do not work as they inherently need
asymptotic free states \cite{kan} and in a mesoscopic set up these asymptotic states naturally can not be in a linear superposition.
Axiomatic quantum mechanics has to achieve this rather artificially which create some limitations.

Once we know $\rho_{lpd}$ we can use it to find other relevant quantities like injectance that we find in axiomatic
framework of quantum mechanics and so we are not loosing anything but only gaining some extra objects by this assertion.
At any point we can go to the closed system
by making the leads $\gamma$ and $\alpha$
very weakly connected to the system and recover all that we get by solving the Schrodinger equation for the closed system.
Besides it may not be always necessary to pinch off the leads physically as von Neumann quantum mechanics give us a method to infer about the
closed system from the scattering states as has been worked out in chapter 4 of Ref \cite{kan}.

With such a local state one can obviously ask the question if it has anything to do with quantum measurements.
Buttiker et al tried to obtain a direct relation between
conductance (a measurable quantity) formulas for mesoscopic systems and local partial density of states (LPDOS).
They could somewhat do so in the semi-classical regime but it did not work for quantum regimes
where $\rho_{lpd}$ can become negative.
So the notion of a local-state and local time in the quantum regime remained unsolved.
A local time, if it
exists in quantum regime, may shed light on how general relativity and quantum mechanics can be reconciled.
These are the two issues, that we will probe in this work.
Namely, how the local time reconciles quantum mechanics and relativity and how the local state help solve the
quantum measurement problem.
The motivation is due to the fact that we could prove \cite{kan} that negativity of $\rho_{lpd}$ is physical
and corresponds to a wave-packet traveling back in time. Since a wave-packet can also carry information or signal,
this raises some philosophical issues. Hence a more down to earth experimental verification will be of great
help and that is also what will be given in this work.

We will first re-interpret Landauer-Buttiker conductance formulas to show that it works because of the reality of local
state and a local time in a quantum system. We will then give examples to show that what ever works in semi-classical
regimes, work much better in the quantum regime. We will also show why Buttiker et al did not get it in
the examples they considered.

\section{Re-interpreting Landauer-Buttiker formalism}

{\bf Landauers approach through DOS vs our approach of using freedom of normalization:}
\begin{figure}
\centering
\includegraphics[scale=0.4]{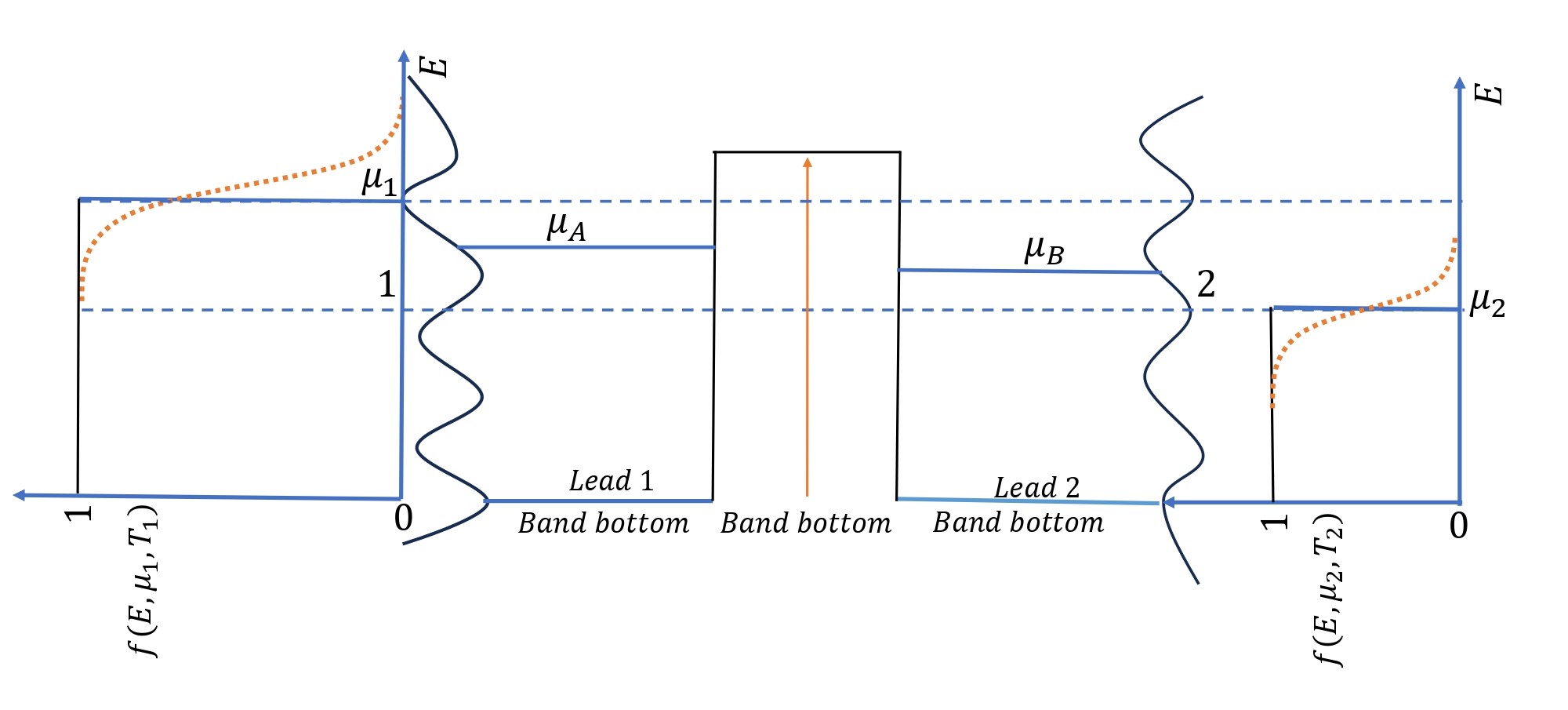}
\captionsetup{labelformat=empty}
\caption{Fig. 1. A 1D quantum scatterer is sandwiched between two classical
reservoirs. The source reservoir is to the left and
the sink reservoir is to the right of the two wavy lines in the figure. The scatterer in between is shown in the form of a finite square
barrier, whose transmission amplitude is discussed in almost all text books of quantum mechanics. The horizontal axis is the
1D coordinate axis (say $x$-axis) and the vertical axis is the energy axis.
The origin of the energy axis
is the band bottom of the conduction band making the sample and the height of the square barrier is shown by an arrow.
Such a 1D square barrier with leads can be practically made from the GaAs-AlGaAs interface by creating a narrow
confinement potential in the lateral $y$ and $z$ directions.
Since the reservoirs are classical, they have a statistically large
number of electrons and the system in between has only quantum states that can transmit electrons from the left
reservoir to the right. The two
reservoirs can have different chemical potentials and different temperatures that can be seen from the Fermi
distributions generally shown inside the two reservoirs.}
\end{figure}
Landauer's original argument
starts from the diagram in Fig. 1 and asserts that it represents a finite physical system where
a) the reservoirs are classical and the
source reservoir injects a statistically large number of electrons
into the left lead that constitute a current for which b) there is a
density of states (DOS) given by $\frac{1}{hv}$.
Assertion b) is troublesome and will be discussed in detail.
There can be no contradiction between reality and assertion a) and allows us to use the axioms of statistical mechanics.
In Ref. \cite{kan}, we have used the axioms of statistical mechanics
and current conservation across a junction
to show that the current through the system at a given incident energy $E$
is $\frac{e_0}{h}|t(E)|^2$
where $t(E)$ is the quantum mechanical transmission amplitude for the rectangular barrier as shown in Fig. 1.
In this approach we do not need to invoke the idea of a DOS to derive the current.
Instead we use the freedom of choosing a normalization constant that conserves current across a junction
with an electrostatic potential, and this normalization constant can be interpreted as a DOS
that accommodate the electrons.
In Eq. 12 we substantiate the original arguments of Landauer, who on the other hand
assumed that this current is carried by electrons in quantum states for which there is a well defined DOS.
By understanding the simple 1D picture in Fig. 1 we can generalize to realistic mesoscopic systems because of the
decoupling of transverse modes in the leads \cite{dat}.
The rectangular
barrier and 1D leads can be replaced by realistic potentials for which the calculation of
transmission probability will be complicated and one can proceed with this simplest case of
a rectangular barrier in mind \cite{dat}.

{\bf Unanswered questions in Landauer's phenomenological approach:}

Many different experiments show that the Landauer formalism and its diverse extensions
to real systems with
multiple leads, finite temperature, many body effects, non-linearities, etc,
remain valid even for very small lead lengths,
as we will discuss.
Hence the assertion b) needs to be scrutinized because $\frac{1}{hv}$ is the DOS for an infinite 1D line for the $+k$ modes only.
The lead connected to the left reservoir is clearly not that. Let us first see how we treat a scattering problem in quantum
mechanics. Therein scattering cross section is defined between asymptotic states. This assumes that the quantum system
extends from $x=-\infty$ to $x=\infty$. The initial state correspond to a wave-packet at $x=-\infty$ and also $t=-\infty$,
when the scatterer is absent. Hence the incident wave-packet actually see an infinite 1D line with translational symmetry
for which momentum and energy are for good quantum states. These states also have wave-vector $+k$ and DOS $\frac{1}{hv}$.
The scatterer is switched on at a latter time wherein the wave-packet goes through a transient period for which there
is no exact theory although there are many approximated ones. For solving fundamental problems one has to ask
what is an approximation to something whose exact behavior is unknown. For example, there is no theory that can
predict transition between bi-stable states or pattern in which super-cooled liquids freeze.
These events happen randomly for which we can develop a probabilistic theory that does not apply to mesoscopic systems
that often do not allow any ensemble averaging.
However, by the time the wave-packet goes to $+\infty$ we also get $t=+\infty$, steady state sets in and the
wave-packet can be written
in terms of the solution of time independent Schrodinger equation. This clearly cannot be adopted to describe the
situation in Fig. 1. In case of Fig. 1, the left lead is finite, there is never a time when the scatterer is absent,
the scattering happen within a finite time, the quantum system does not extend from $x=-\infty$ to $x=+\infty$.
The left reservoir injects an initial right moving wave-packet into the left lead for which the Fourier
components are very appropriately
$e^{ikx}$, but that is quite different from an asymptotic state for various reasons.
First is that $k$ is a wave-vector that has nothing to do with momentum state because the scatterer is never
absent and there is never a translational symmetry. From left reservoir to right reservoir, energy is conserved
but that is different from an eigen-state of the Hamiltonian. The mode $k$ can vary continuously, not because it is an
element of a continuous spectrum but rather because energy inside a classical reservoir, say the cathode of
a battery, can vary continuously. This classical energy of the electron interacts with the eigen-states
of the Hamiltonian operator in the system that Landauer describes as a quantum mechanical
scattering phenomenon. Besides the wave-function in the left lead is not $e^{ikx}$ but is
$e^{ikx} + re^{-ikx}$ and it is this reflection term that make a standing wave
all the way up to $x=-\infty$ and that is
different from an asymptotic state. The DOS for such an infinite system correspond to
energy states, are to be determined from the Greens function, and is different from
$\frac{1}{hv}$, and in fact this difference is very troublesome when we try to extend
this theory of scattering to three dimensions.

Finite quantum systems can in practice have asymptotic states and
a well defined DOS,
that is given by the resolvent and Greens function where the exact
scattering potential has to be taken into account. This is related to the dwell time to be discussed below (see Eq. 11),
where the integration limits is not for the region of length $l$ of the barrier, but has to
be over the entire region between the junctions with the reservoirs.
Such finite quantum systems will not have momentum states but
can have energy states, however, their DOS is not $\frac{1}{hv}$ as Landauer assumed.
For the system in Fig. 1, between the reservoirs, there is no dissipation and so energy will be conserved, energy states may be taken as
good quantum states, but that will contradict how Pauli exclusion principle is applied as will be discussed.
This is essentially because $\pm k$ modes are not momentum states and rather belong to the same energy state.
Besides, the term "number of states"
is sometimes misleading as one cannot count states in the continuum of states. This contradiction of terms
is not a serious matter for large systems because of a mathematically rigorous prescription of
regularization that can be applied to real systems but that prescription is not adaptable to small mesoscopic systems.
This is essentially because regularization scheme applies to continuous spectrum of practically infinite systems,
which is clearly not the case that Landauer had in mind where the total length between the reservoirs is comparable to
elastic mean free path and de Broglie wave-length, and the incident electrons are coming from a classical reservoir with continuously
varying classical energy.
In formal quantum mechanics only infinite systems have a continuous spectrum implying that states are uncountable. Therefore we can
start from some finite system for which infinite limits are well defined, to get a
DOS that is defined as the number of states per unit energy per unit length, which is then
regularized to give measure or volume of states of the full
system at an energy divided by the volume of the system to {\bf mean} that at a given point  or
cross section. Mesoscopic systems being finite by definition, do not allow this rigorous prescription
of regularization and yet electrons coming from the classical reservoir has a continuously varying incident energy.

Ideas we use in condensed matter system cannot be applied here because we know that in condensed
matter systems an infinite periodic potential has real eigen-energies or eigen-states
of the Hamiltonian operator that are extended. Also an infinite random potential has
real eigen-energies that are localized.
Mesoscopic phenomenon arise from finiteness of the system.

{\bf Semi-classical approach:}
In condensed matter systems, WKB approximation works well that tells us there
can be a local momentum corresponding to the
classical momentum. If we assume these features then we are still in the condensed matter regime and not
accessing the mesoscopic regime.
The prescription followed by S. Datta \cite{dat} based on
use of Dirac delta function is ad hoc. Dirac delta function is good for solving linear differential
equations but it is difficult to translate what a Dirac delta function means for finite
quantum systems in nature and whether it helps us bypass
the standard mathematical rules of regularization by introducing a complex self-energy parameter for a finite system.
For the system size tending to infinity, there is a very close connection between the so called von Neumann's approach
through Hilbert space and
Dirac's approach through bra-ket notation. One being rigorously mathematical and the other being more intuitive,
with the caution that the intuitive approach may be very helpful but can overlook errors.
Dirac's approach looses its effectiveness unless space is discritized and the Hamiltonian is represented as a matrix.
Thus the self-adjoint Hamiltonian operator is reduced to a Hermitian matrix to give Heisenberg's matrix mechanics,
and then the continuum limit can be imposed through a hopping parameter without following the mathematical rigor
of taking limits.
If one is ready to accept some parameter fitting then this method adopted
by S. Datta \cite{dat} is good for numerical calculations
and hopefully that is adequate in the semi-classical regime.
One may state that such ad-hoc approaches cannot reconcile the theories of quantum mechanics and relativity.
One may ask what role does the hopping parameter
or the self energy parameter play in general theory of relativity
and how do they transform.
Besides as it has been shown \cite{kan} that one cannot define a momentum operator for a general mesoscopic system
in the quantum regime,
canonical quantization will not work and von Neumann's wave-mechanics is essential.

{\bf Our fix for what is a state in a mesoscopic lead for which we write a DOS?} For this note the fact that given a problem,
von Neumann quantum mechanics first require us to determine the Hilbert space.
The Hilbert space is determined by specifying the basis-states and the operators that transform the basis states.
We can use the eigen-states of some well defined operator as the basis-states.
Since there is only one Hilbert space (that is isomorphic to the complex plane), any complete orthonormal basis may work, only restrictions can come from
boundary conditions or physical constraints or dimensionality of the sample, etc, that restrict the theory to a subspace.

Quantum mechanics start with the fact that nature generates a wave-packet that evolves in time,
for example, the classical reservoirs inject a pulse of classical electrons into the system.
Inside the mesoscopic system, the pulse has to treated as a wave-packet using
time dependent Schrodinger equation (TDSE) and quantum mechanics
start with this wave-packet. The wave-packet
can be decomposed into partial waves that satisfy time independent Schrodinger equation
(TISE) and quantum mechanics tells us how to relate the two dynamics using the probabilistic
interpretation and uncertainty principle. Essentially many events of pulse propagation averages to
give the scattering probabilities and all other physical effects that one can get from the
steady beam scattering. A steady beam or a stationary state can be only viewed as a limiting
situation of uncertainty in momentum going to zero and taking this limit has to be treated with
extreme mathematical rigorousness. This cannot be done by assuming that a Kronecker delta smears into
a Dirac delta function as is often done \cite{dat} (see page 97 therein).

Finally, a wave-packet (also a Hilbert space element at an instant
of time) is what starts from an initial condition
and propagates in real space and has a time dependent equation of motion. The wave-packet can be
mathematically constructed from wavelets that
have wave-vectors that vary continuously.
While discrete objects can be counted, continuum of wave-vectors has a density.
We can completely ignore the notion of DOS in quantum mechanics and
instead arrive at a local time recorded by a physical clock (a Larmor clock) \cite{lar}
in the form of a spin of an electron in the wave-packet.
The physical clock gives a correspondence between quantum mechanical DOS and angular displacement
of the precessing dipole of the spin.
The angular displacement of the dipole
divided by the angular velocity of classical Larmor precession give the local time and further divided by $h$
give DOS which is completely consistent with the DOS we define quantum mechanically \cite{lar}.
Thus DOS can be interpreted as how many (or volume of) wave-vector values are needed to make a topological circle in a complex plane, say
of the complex transmission amplitude $t$ ($Im(t)$ versus $Re(t)$) \cite{kan},
as will be explained below.
Thus the idea of a quantum number is replaced by winding number in this complex plane.
Thus there is something
like a local quantum number at the point $\textbf r$
which is more appropriately a fractional winding number
$\delta \theta_r$ or $\delta \theta_t$
for which there is a local DOS (called local partial density of states or LPDOS)
\begin{equation}
\rho_{lpd}^r=-\frac{1}{2 \pi e_0}\frac{\delta \theta_r}{\delta U(\textbf r)} \;\;\mbox{and}\;\;
\rho_{lpd}^t=-\frac{1}{2 \pi e_0}\frac{\delta \theta_t}{\delta U(\textbf r)}
\end{equation}
at a point $\textbf r$ inside the scatterer,
for the reflected and the transmitted electrons, respectively.
Here $\theta_r$ and $\theta_t$ are the phases
of the complex reflection amplitude $r$ and transmission amplitude $t$, respectively, that change with $k$ and hence $U(\textbf r)$,
$U(\textbf r)$ being the electro-static potential at the
point $\textbf r$ and the derivatives are functional derivatives.
Here $e_0$ is the charge of the particles.
This local DOS in Eq. 1 is specifically called local partial density
of states (LPDOS) as it holds for only reflected electrons or only transmitted electrons.
We have shown that this can
be justified through physical clocks as well as functional analysis, without relying on operators in quantum
mechanics and can be seen in more detail in \cite{kan}.
The objective reality of LPDOS can
be established from functional analysis without demanding quantum states. It is consistent with
quantum mechanics, but it cannot be found within the axiomatic framework of quantum mechanics \cite{kan}.
The winding numbers gradually evolve into discrete quantum numbers of appropriate
operators, as we pinch off the system and isolate it
from external incidence from a classical reservoir.
At low energy $U(\textbf r)$
is classical with no quantum corrections while only the particles are fully subject to
quantum mechanics.
However, in appropriate conditions, $U(\textbf r)$ can exhibit quantum fluctuations for which there is a well established theory.

In quantum mechanics "states" is an ill defined concept and science deals with such ill defined concepts in terms of properties
or qualities for which we use an appropriate adjective to get a qualified state.
One can cite other examples like we do not understand what is time but we can define diffusion time,
or first passage time etc. We do not understand causality but we can define space-time-causality, or interventionalist's-causality.
Our local partial density
of states (LPDOS) defined in Eq. 1, are just states and not to be hyphenated with any operator but consistent with appropriate states of a quantum operator,
depending on the problem at hand.
This is further elaborated as we proceed.

To find the LPDOS at a point $\textbf r$ in the right lead (say) we have to note that $\theta_t$
is the same at all points on the right lead
and the phase shift to be considered at a distance
of $l$ from the scatterer is
\begin{equation}
\theta_t+kl
\end{equation}
were $l$ is the length between the end point of the square barrier potential and the point $\textbf r$ of the right lead.
Electrostatic potential $U$ in the leads is given by the bottom of the conduction band and is constant.
Therefore, in the leads,
\begin{equation}
k=\sqrt{\frac{2m_0}{\hbar^2}(E-U)}
\end{equation}
Comparing Eqs. 1, 2 and 3, we can say that to talk about total states in the right lead available for the transmitted electrons,
we have to apply an infinitesimal $U(\textbf r)$ at the point $\textbf r$ on the
right lead (see Eq. 1) and then integrate over $\textbf r$ in the right lead.
Hence,
in the right lead, from Eq. 3, one gets
\begin{equation}
\frac{d}{dE}\equiv -\frac{1}{e_0}\underset{\mbox{right lead}}{\int} d\textbf r \frac{\delta}{\delta U(\textbf r)}
\end{equation}
leading to the fact that electrons access a
number of states in the right lead at a point and at an energy $E$ given by $\frac{1}{hv}$.
This can be seen as follows.
\begin{eqnarray}
\rho_{pd}^{rl}=
\int d\textbf r \rho_{lpd}^{rl}=
-\frac{1}{2 \pi e_0}\int d\textbf r \frac{\delta (\theta_t+kl)}{\delta U(\textbf r)}=
-\frac{1}{2 \pi e_0}\int d\textbf r \frac{\delta (\theta_t)}{\delta U(\textbf r)}
-\frac{1}{2 \pi e_0}\int d\textbf r \frac{\delta (+kl)}{\delta U(\textbf r)}=
-\frac{1}{2 \pi e_0}\int d\textbf r \frac{\delta (+kl)}{\delta U(\textbf r)}
\end{eqnarray}
Here $\rho_{lpd}^{rl}$ refers to local partial density of states at any point (hence local) available to electrons that are
right going in right lead (hence partial)
and $\rho_{pd}^{rl}$ refers to that integrated over the entire finite length of the lead.
Note that $\theta_t$ is independent of a vanishingly small
$U(\textbf r)$ in the lead acting on the partial $+k$ mode,
that is $\frac{\delta (\theta_t)}{\delta U(\textbf r)}\bigg\vert_{\textbf r \in \mbox{(right lead})}=0$.
This $U(\textbf r)$ is not meant to reflect a transmitted electron but only meant to
give a derivative (like a tangent to a curve does not need to increment the curve).
Also note that Eq. 4 says that increasing the incident energy for a fixed $U$ in the lead, is equivalent
to decreasing the $U$ in the lead for a fixed incident energy, which is obvious from Eq 3.
Therefore from Eqs. 4 and 5 and noting the fact that $k$ is constant for all $\textbf r$ in the lead
($k$ will however have a gradient with $U(\textbf r)$ for $E$ constant or will have a gradient with E for $U$
constant),
\begin{eqnarray}
\rho_{pd}^{rl}=
\frac{1}{2\pi} \frac{d(+kl)}{dE} =
\frac{1}{2\pi} \frac{d(+kl)}{d(+k)} \frac{d(+k)}{dE_{(+k)}}
=\frac{1}{2\pi} \frac{d(+kl)}{d(+k)} \frac{d(k)}{dE_{(k)}}
=\frac{1}{2\pi} \frac{d(+kl)}{d(+k)} \frac{1}{\frac{dE}{dk}}
=\frac{1}{2\pi} \frac{l}{\frac{d}{dk} \frac{\hbar^2 k^2}{2m_0}} \nonumber \\
= \frac{2m_0l}{2 \pi \hbar^2} \frac{1}{2k}
= \frac{m_0l}{2\pi\hbar^2 k} = \frac{l}{h} \frac{1}{\frac{\hbar k}{m_0}} = \frac{l}{hv}
\end{eqnarray}
Usually in text books there is an extra factor 2 to account for $-k$ but we are only considering $+k$ partial modes to which we will
locally apply Pauli exclusion principle as justified below.
The transmitted electrons are in the $+k$ partial mode and therefore,
\begin{equation}
\rho_{pd}^{rl}= \frac{l}{hv}
\end{equation}
Since LPDOS physically (not just meant) add up to give PDOS
\begin{equation}
\rho_{lpd}^{rl}= \frac{1}{hv}
\end{equation}
Thus from Eq. 8, we have local partial states at a point on the right lead with a density given by
$\rho_{lpd}^{rl}=\frac{1}{hv}$
and for the extended region of the right lead, we get partial states with a density given by
$\rho_{pd}^{rl}=\frac{l}{hv}$ (from Eq. 7) and both are equally physical because LPDOS is a physical object that is coordinate dependent
in general although not explicitly in the leads,
that add up to give PDOS. This is thus different from DOS for the states of the full system that we get from axiomatic quantum mechanics
that can be mathematically divided by the volume of the system to mean something.

{\bf Pauli exclusion in the leads} In the mesoscopic leads,
the Landauer-Buttiker formalism assumes Pauli exclusion principle for each wave-vector separately for
transmitted and reflected electrons that definitely does not define a momentum-state.
Even if we assume energy-states then incident electron and reflected electron belong to the
same state and we cannot apply Pauli exclusion principle to both of them independently.
At zero temperature, all incident modes are occupied and so nothing can be reflected.
We can justify this assumption in the formalism as done below from LPDOS.
Let us consider the reflected electrons first. We understand that
$\rho_{lpd}^{ll}$ is
LPDOS of reflected electrons in $-k$ mode, in the left lead and is due to
a time spent \cite{kan} by the reflected particle at the coordinate $\textbf r$
that is given by
\begin{equation}
\Delta \tau_{lpt}^{ll}=-\frac{\hbar}{e_0}\frac{\delta \theta_r}{\delta U(\textbf r)}
\end{equation}
This time is consistent with the mesoscopic Larmor clock which in turn is consistent with Lorentz's transformation as
\begin{eqnarray}
U\rightarrow \gamma U\;\;\mbox{and}\;\;
\gamma=\frac{1}{\sqrt{1-v^2/c^2}}
\end{eqnarray}
and gives proper time (Larmor clock is body frame clock)
correctly that is less than the coordinate time difference between two classical points in the classical reservoirs.
Hence the local state at the point $\textbf r$ in the lead is consistent with quantum mechanics as well as
Lorentz's transformation and therefore by the
statement of spin statistics theorem, we can apply Pauli exclusion principle to the local state defined by a winding number.

Similarly, at every point $\textbf r$ in the right lead, we can
apply Pauli exclusion principle to the mode $+k$.
So the  LPDOS for $+k$ mode in the right lead turns out to be the same at all points on the right lead,
is $\frac{1}{hv}$, and it is subject to Pauli exclusion principle.
However, for $\textbf r$ inside the system with a constant potential
barrier, $\frac{d\theta_t}{dE}\not\equiv \frac{1}{e_0}\frac{\delta}{\delta U}\theta_t$
because of linear superposition of $\pm k$-modes
and there is no decoupling of the $\pm k$ modes due to multiple Feynman paths.
For example, $\frac{1}{e_0}\frac{\delta}{\delta U(\textbf r)}\theta_t$
will be $\textbf r$ dependent, depending on how close we are to the edge of the constant potential barrier.
In the leads, this superposition is destroyed by the classical reservoirs \cite{kan}.
For a non-rectangular barrier, say for a wavy potential barrier, inside the system there is no identifiable
$\pm k$ states,
$\frac{d\theta_t}{dE}\not\equiv \frac{1}{e_0}\frac{\delta}{\delta U}\theta_t$,
but the leads can again be treated exactly like that of the leads of a rectangular barrier.

Every point inside the sample can thus be assigned a
coordinate time (obtained from Eqs. 1 and 10 with $v\rightarrow 0$) that is in the laboratory frame and from this frame the body clock runs slow for $v \neq 0$.
In analogy, in a typical Stern-Gerlach set up where spins
precess, every point inside the system can have three spatial coordinates
due to three perpendicular spatial axes
(the classical magnetic field can be
oriented in three perpendicular directions)
while for each independent spatial direction,
the two spin degrees of freedom define an internal orthogonality in the body frame.
Following Eqs. 1 and 10, one can assign three spatial coordinates
and one time coordinate to every point in the sample and there is an additional internal body frame clock that records
a proper time that is consistent with relativity.
This is also true for the classical Larmor clock
for which one can see \cite{ltpt}.

Eqs. 7, 8, 9 and 10 are consistent with a dwell time in a section of length $l$
in the leads that can be defined within the axiomatic framework
of quantum mechanics. For the electrons incident from the left reservoir into the left
lead we can write the wave-functions as $\psi(k,x)=e^{ikx}$ and hence the dwell time as
\begin{equation}
\frac{1}{v}\int_0^l dx|\psi(k,x)|^2=\frac{l}{v}
\end{equation}
and divided by $h$ a DOS. Here $l$ can be arbitrarily small or large.
The dwell time can be found within the axiomatic framework of quantum mechanics as in a general
situation (including the situation in a lead where superposition of states is destroyed by a reservoir) $\psi(k,x)$ inside
the system does not have any information of from where the electrons come and where they go, but not LPDOS.

{\bf Outline:} While the general view is that quantum mechanical laws are the fundamental laws of nature and classicality emerges,
we adopt the view that a quantum event can only be defined between two classical events in
the reservoirs and the quantum laws can
coexist with classical laws without any contradiction.
This has some obvious but significant consequences and solves some long standing problems in physics.
First is that theory of relativity and quantum mechanics can co-exist consistently and there is no immediate need to
quantize gravity, besides the fact that $U(\textbf r)$ has well known quantum behavior.
That is because now in quantum mechanics, like in theory of relativity, we can relate time intervals
to signal propagation time. It was shown in ref. \cite{kan} that this signal
is carried by a propagating localized undispersed quantum mechanical wave packet. We can clearly obtain a coordinate time and a proper time
within the framework of quantum mechanics.
The points inside the classical reservoirs have local clocks giving their respective local times or coordinate time
that is consistent with mass distribution in synchronizing the classical local clocks.
A signal can propagate from one reservoir to another reservoir according to the laws of quantum mechanics and the
time delay defines a proper time that can dilate and contract
and can be different from local times clocked by the classical clocks inside the reservoirs.
The quantum system gets a space-time formulation in the sense that every
point inside the quantum system has three spatial coordinates and a time coordinate.
This time coordinate or coordinate time ($v \rightarrow 0$
in Eq. 10) is consistent with proper time that will be recorded
by the body clock of a particle or wave-packet traversing the points and that in turn defines an un-hyphenated local quantum state
that does not need a Hilbert space.
Thus we get grand unification of quantum and classical laws at low energies
rather than at high energies.

One can present three very
strong evidences in favor of this view. First is that this is supported by the enormously successful Landauer-Buttiker formalism as will
be shown in sub-section 2.1. Second is that it leads to a natural understanding of what happens in a measurement on a quantum system, that
is, solves the measurement problem in quantum mechanics as will be shown in sub-section 2.2.
Third is that it leads to a definition of time in a quantum system that is fully consistent with STR and GTR,
and this time does not crop up in the von Neumann's axiomatic framework of quantum mechanics and has been missed so far,
but it exists in nature as will be shown in Fig. 5.
These points will be exemplified through calculations below.
This also implies that unlike in von Neumann quantum mechanics or any other version, we do not have a quantum
dynamics extending from $-\infty$ to $\infty$ and that can naturally remove certain divergences.

\subsection{Application to Landauer-Buttiker conductance formulas}

The above mentioned idea of exclusion principle applied to local partial states
will be now used to re-describe Landauer-Buttiker formalism.
We will apply exclusion principle to local partial modes at zero temperature
as well as to finite temperatures.
It will be thus clear that Fermi-Dirac statistics applies to local
partial $k$ modes and that finds strong experimental support in the overwhelmingly successful
Landauer-Buttiker conductance formulas and is explained in detail below.
There are no states and no Fermi-Dirac distribution function in the reservoirs unlike what is
presented in Fig. 1. Yet as a standard practice,
one typically draws a Fermi-Dirac distribution function in the reservoirs as shown in Fig. 1,
and as can be seen in the book by S. Datta \cite{dat}.
This is un-necessary or may be a matter of {\bf convenience}, but could also be related to a typical mindset that
at very low temperatures the reservoirs become quantum mechanical.
The general belief is that quantum mechanics is fundamental and classicality emerges.
As a consequence of this general belief,
one may think that at zero temperature the
whole universe will become quantum mechanical with the caveat or circular logic that zero temperature
can never be reached.
In this general belief system one draws a Fermi-Dirac distribution inside a reservoir.
Mesoscopic experiments are done at extremely low but finite temperatures such that
$k_BT$ is less than sub-band spacing in the leads and one should never consider the reservoirs
to be at zero temperature.
A classical reservoir can have a chemical potential and temperature which are just parameters and
not related to a Fermi-Dirac distribution of electrons.
Fermi-Dirac distribution can be however, considered once the electrons enter the leads and the system
that are quantum mechanical.
In the region between the reservoirs we have a scattering problem that is described in general
in the framework of von-Neumann quantum mechanics \cite{kan} (see ch 4 therein) that can be easily adapted
for the 1D case of Fig. 1.
In the Landauer's approach, the mathematics
of quantum mechanics connects to currents in nature through $ne_0v$ where $n$ is the number (countable) of carriers which in this case is
electrons, $e_0$ is the charge of an electron and $v$ is its velocity.

{\bf Two probes:} The left reservoir injects a current into the
mesoscopic sample for which we have discussed that the available states locally at any point for $+k$
(see Eqs. 8 and 9) is say $\rho^{ll}_{lpd}(+k)=\frac{1}{hv}$, where $v=\frac{\hbar k}{m_0}$,
$m_0$ being the mass of an electron. For right moving electrons $k$ is positive, while negative for
left moving electrons. So total injected current in the left lead, at a point close to the reservoir,
between $\mu_1$ and $\mu_2$,
that is in the energy range $dE=(\mu_1-\mu_2)$ will be
\begin{equation}
dJ_{in}=(dn)e_0v=\frac{dn}{dE}e_0v(dE)=\rho^{ll}_{lpd}(+k)e_0v(dE)=\frac{1}{hv}e_0v(dE)=\frac{e_0}{h}(\mu_1 - \mu_2)
\end{equation}
Therefore the differential current that pass through the sample to reach the sink reservoir is
\begin{equation}
dJ=\frac{e_0}{h}|t(E)|^2(\mu_1 - \mu_2)
\end{equation}
If there is a current between a source and a sink then one can easily define a conductance as will be done below.
In a typical two probe conductance measurement wherein the voltage difference is measured between the two
classical reservoirs, the conductance will be
\begin{equation}
G_{two-probe}^{zero-temperature}=\frac{\frac{e_0}{h}|t(E)|^2(\mu_1 - \mu_2)}{\frac{\mu_1 - \mu_2}{e_0}}=\frac{e_0^2}{h} |t(E)|^2
\end{equation}
If we consider spin degeneracy for positive $k$ then DOS will be $\frac{2}{hv}$ and this formula becomes
\begin{equation}
G_{two-probe}^{zero-temperature}=\frac{\frac{2e_0}{h}|t(E)|^2(\mu_1 - \mu_2)}{\frac{\mu_1 - \mu_2}{e_0}}=\frac{2e_0^2}{h} |t(E)|^2
\end{equation}

Thus we have adopted the fact that
inside the system, when $k_BT$ is less than sub-band spacing due to lateral confinement, we can apply
zero temperature quantum mechanics.
At zero temperature, all states
in the conduction band is filled below $\mu_2$ and are also equilibrated which can also be seen
as the principle of detailed balance as will be explained below.
The reservoir has no Fermi energy but even classical reservoirs have a chemical potential and
temperature. When the electron enters the lead then it gets a chemical potential which is also
the Fermi energy.

We will now proceed to account for finite temperature effects, voltage differences and role of multiple leads.
All these formulas has been rigorously shown to agree with experiments [see \cite{dat}]
in spite of the severe conceptual problems
discussed above.
We remove these conceptual problems and thereby show that this naturally solves other problems in physics.

{\bf Finite temperature : }
Now let us see what may happen at finite temperature when states above $\mu_1$ and $\mu_2$
has a finite probability to be occupied.
The left lead will be in equilibrium with the left reservoir. Say there are
$\frac{d n}{d E}$ number of relevant modes at any (all) point
in the left lead and not all of them are occupied. Number of electrons at a given energy $E$, at any point in the left lead will be
given by the number of occupied modes, that is
$ \frac{d n}{d E} f(E,\mu_1) $
where
$f(E,\mu_1)$ is the Fermi-Dirac distribution function.
Similarly,
number of electrons at any point on the right lead is
$\frac{d n}{d E} f(E,\mu_2)$.

So number of candidate electrons on the left of the scatterer that wants to cross to the right is
\begin{equation}
d n \bigg(f(E,\mu_1) - f(E,\mu_2)\bigg)
=\frac{d n}{dE} \bigg(f(E,\mu_1) - f(E,\mu_2)\bigg) dE
\end{equation}
Occupied $+k$ modes in left lead contribute positively, while occupied $+k$ modes in right lead are subtracted to get Eq. 16.
For the $+k$ modes only, the relevant DOS is $\frac{dn}{dE}=\frac{1}{hv}$ in both leads.
Number of them that actually cross to the right is
$$\frac{d n}{d E} \bigg(f(E,\mu_1) - f(E,\mu_2)\bigg) |t(E)|^2 dE$$
Current $ne_0v$ through the system will therefore be
\begin{equation}
\frac{d n}{d E} \bigg(f(E,\mu_1) - f(E,\mu_2)\bigg) |t(E)|^2 e_0v dE
\end{equation}
$$ = \frac{1}{hv} \bigg(f(E,\mu_1) - f(E,\mu_2)\bigg) |t(E)|^2 e_0v  dE$$
$$ = \frac{e_0}{h} \bigg(f(E,\mu_1) - f(E,\mu_2)\bigg) |t(E)|^2 dE $$
$$ = \frac{e_0}{h} \frac{\partial f}{\partial \mu} d\mu |t(E)|^2 dE = \frac{e_0}{h} (-\frac{\partial f}{\partial E}) (\mu_1 - \mu_2)|t(E)|^2 dE
\;\;\mbox{as}\;\; \frac{\partial f}{\partial \mu} = - \frac{\partial f}{\partial E}$$
Total current at finite temperature will be
\begin{equation}
J_{two-probe}^{finite-temperature} = \int_{0}^{\infty} \frac{e_0}{h} (- \frac{\partial f}{\partial E}) |t(E)|^2 (\mu_1 - \mu_2) dE
\end{equation}
$$ J_{two-probe}^{finite-temperature} = \frac{e_0}{h} (\mu_1 - \mu_2) \int_{0}^{\infty} dE (-\frac{\partial f}{\partial E}) |t(E)|^2$$
For the two probe conductance one has to measure the voltage difference between the two classical reservoirs, which again turns out to be
$$G_{two-probe}^{finite-temperature} =
\frac{J_{two-probe}^{finite-temperature}}{\frac{\mu_1 - \mu_2}{e_0}} = \frac{e_0^2}{h}\int_{0}^{\infty} dE (-\frac{\partial f}{\partial E}) |t(E)|^2 $$
Let us take the zero temperature limit wherein
$f(E) \rightarrow \theta(E_f - E)$
and $(-\frac{\partial f}{\partial E}) \rightarrow \delta(E - E_f)$ to get
\begin{equation}
G_{two-probe}^{zero-temperature} =\frac{e_0^2}{h} |t(E_f)|^2
\end{equation}
Since the Fermi energy or the incident energy is determined by the left reservoir we may replace $E_f$ by $E$ to get
\begin{equation}
G_{two-probe}^{zero-temperature} =\frac{e_0^2}{h} |t(E)|^2
\end{equation}
which is same as Eq. 14.

Assuming an exact LPDOS of $\frac{1}{hv}$ in the finite leads
play a further role when we discuss charge neutrality in the leads as will be presented below.
It assumes that there are wave-vectors $+k$ and $-k$ to which we can apply Pauli
exclusion principle separately. The $\pm k$ wave-vectors naturally come from the wavelets making the wave-packet which is the
central object of study in quantum mechanics. That $-k$ states are occupied only if there is reflection
(as will be done below) from the scatterer
etc, but to apply Pauli exclusion principle for the $\pm k$ modes separately without questioning whether they define quantum states is ad hoc.
While developing a conductance formula it is not convenient to talk of energy states, and although one may argue that
energy is conserved between the reservoirs,
$\pm k$ belong to the same
quantum state (energy state), which would mean that if $+k$ state is occupied then $-k$ state cannot be occupied.
In fact that the current through a system is given by the transmission probability through the system was known
much before Landauer and it was thought, that happens only when we have sparsely filled band or very low
transmission probability such that multiple electrons competing for a state is negligible.
One may see \cite{cbd}
or can also see the more recent work of \cite{rtle}.
Surprise came with the success of Landauer approach for densely filled bands, transmission probability approaching the
maximum limit etc.
A lot of space has been given in S. Datta's book to discuss this matter but a confusion can only be expressed in
a confusing manner as can be seen on page 93 of \cite{dat}.

{\bf Towards a four probe Landauer conductance formula at zero temperature:}
Note that the central object in quantum mechanics that is a wave-packet can be decomposed
into wavelets and that naturally give four different wavelets. They are $+k$ mode moving forward in time, $+k$ mode moving backward
in time, $-k$ mode moving forward in time and $-k$ mode moving backward in time. In this section we are only talking about $\pm k$ modes
in a wave-packet moving forward in time because the initial wave-packet is moving forward
in time. A $+k$ mode moving in forward time is physically indistinguishable from a $-k$ mode moving
in backward time and to keep the
arguments simple we only consider forward moving modes. If backward moving modes manifests for the
given initial condition of the wave-packet (for example it does
in the vicinity of a Fano resonance when the wave-packet that was initially moving forward
in time starts moving backwards
in time \cite{kan}) the arguments can be accordingly adjusted easily.
Once again by zero temperature we only mean $k_BT$ is less than the sub-band spacing in the leads (see ch4 in \cite{kan}).
For the two probe conductance, the voltage difference between the two classical reservoirs are considered. For a four probe conductance we need the voltage difference
between the two leads and the leads are quantum in nature, thus leading to complicacies
related to measurements in quantum mechanics. Theoretically we can always calculate a voltage difference between
the two quantum leads based on the countability
of electrons, and so we proceed to calculate the chemical potentials $\mu_A$ of left lead and $\mu_B$ of right lead, at zero temperature.
For a completely opaque barrier we will get the situation $\mu_A=\mu_1$ and $\mu_B=\mu_2$. For a barrier that transmits, we will
get a situation $\mu_1>\mu_A>\mu_B>\mu_2$ as shown in Fig. 1.
First consider the right lead where we only have to consider the $+k$ modes
for which the number of available states at a point
is given by the LPDOS, and for the relevant current carrying energy window the number of electrons is
$\frac{d n}{d E}(\mu_1-\mu_B)$.
Of all these available states, a state will be occupied only if it receives an electron transmitted across the scatterer from the left reservoir.
Therefore the number of occupied local states in right lead, between $\mu_B$ and $\mu_1$ is
\begin{equation}
n^{oc}_{rl}= \frac{d n}{d E} (\mu_1 - \mu_B)|t(E)|^2
\end{equation}
There will be also some unoccupied local states between $\mu_2$ and $\mu_B$ that can be found as follows.
The total number of available local states in this range is
$$ 2 \frac{d n}{dE} (\mu_B - \mu_2)$$
The factor 2 comes because in this range of energy we have to consider right moving $+k$ states as well as left moving $-k$ states.
The $+k$ states end in the right reservoir and $-k$ states originate from the right reservoir.
If we imagine a situation when all these states are unoccupied and then some of the $+k$ states will be filled up by
electrons transmitted across the scatterer from the left.
Which means, of the $ \frac{d n}{dE} (\mu_B - \mu_2) $
right moving states in the right lead, the number of unoccupied ones will be
$$ \frac{d n}{dE} (\mu_B - \mu_2) - \frac{d n}{dE} (\mu_B - \mu_2) |t|^2 $$
All the $-k$ states will be unoccupied because the only way that these states can be occupied is if they receive an electron from the right
reservoir and at zero temperature this cannot happen in the energy interval $\mu_B-\mu_2$.
Any transmitted electron from left reservoir or reflected electron from the right reservoir can only go to a $+k$ state.
Which means all the $ \frac{d n}{dE} (\mu_B - \mu_2) $ left moving
states in this energy range will be unoccupied in the right lead. So the total number of unoccupied states in right lead will be
\begin{equation}
n^{un-oc}_{rl}= 2 \frac{d n}{dE} (\mu_B - \mu_2) - \frac{d n}{dE} (\mu_B - \mu_2) |t(E)|^2
\end{equation}

Let us now consider the left lead.
Number of occupied local states at a point between $\mu_A$ and $\mu_1$
\begin{equation}
n^{oc}_{ll}=\frac{d n}{dE} (\mu_1 - \mu_A)+ \frac{d n}{dE} (\mu_1 - \mu_A)|r(E)|^2= \frac{d n}{dE} (\mu_1 - \mu_A) (1 + |r(E)|^2)
\end{equation}
The first term is for $+k$ modes and all of them are occupied
by electrons coming from left reservoir. The second term is for $-k$ modes that are occupied only if there
is reflection whose probability is $|r(E)|^2$.
This is where one has to question what are the states because $+k$ mode electron is being coherently reflected into a $-k$ mode
and we are applying Pauli exclusion principle to them as if they are separate states.
Since we can apply Pauli exclusion principle to local partial states, this is justified.
Number of unoccupied states in left lead between $\mu_2$ and $\mu_A$ will be
\begin{equation}
n^{un-oc}_{ll}=2 \frac{d n}{dE} (\mu_A - \mu_2) - \frac{d n}{dE} (\mu_A - \mu_2) (1+|r(E)|^2)
\end{equation}
The first term gives the total number of local states at a point. From them if we subtract the occupied ones then we will get the number of unoccupied states.
So the second term can be understood by looking at Eq. 23.
Charge neutrality at a point in the lead requires
\begin{equation}
n^{oc}_{ll}= n^{un-oc}_{ll} \;\;\mbox{and}\;\; n^{oc}_{rl}= n^{un-oc}_{rl}
\end{equation}
Since we are equating occupied states to unoccupied states to get charge neutrality,
again what we mean is that a local partial state can be occupied
by one electron only, at the expense of an un-occupied local partial state.
Solving Eq. 25 one can get
\begin{equation}
\mu_A - \mu_B = (\mu_1 - \mu_2)|r(E)|^2
\end{equation}
If this voltage difference can be measured then we will get a conductance that can be obtained from Eqs. 13 and 26
\begin{equation}
G^{hypothetical}_{four-probe}=
\frac{\frac{e_0}{h}|t(E)|^2(\mu_1 - \mu_2)}{\frac{\mu_A - \mu_B}{e_0}}=
\frac{e_0^2}{h}\frac{|t(E)|^2}{|r(E)|^2}
\end{equation}
It is labeled hypothetical because this conductance is purely theoretical and what will be measured
by attaching the voltage probes to the quantum mechanical leads,
cannot bypass the measurement problem of quantum mechanics. That is analogous to eigen-states of an atom
that can be calculated but cannot be observed because any attempt to observe will disturb the state. This also tells us why a conductivity
of a mesoscopic sample can be calculated from Eq. 27 but cannot be observed.
This also makes it clear that the Hamiltonian of a system cannot tell us the conductance.

The actual four probe measurement revealed that conductance is not symmetric in flux.
This created a lot of confusion because reversing the field was thought of as a change of perspective
with respect to the observer. Field directed into the sample or out of the sample is always with respect
to from where the observer is looking at the sample. In case of the two probe conductance where
the measuring devices are all connected to the classical reservoirs, for a given direction of applied field,
the direction can be considered positive or negative by moving the observer behind or in front of the sample.
Two probe conductance was indeed symmetric in flux, meaning independent of flux direction, and hence
also independent of whether the observer is in front or behind. But since four probe measurement
depend on direction of flux, it was initially thought that the point of view of the observer had something
to do with the nature of collapse of wave-function in a quantum measurement. A faint clue as to
what could be happening is that the asymmetric in flux conductance was however, reproducible,
while collapse of wave-function is supposed to be random.
As to what is actually observed or measured in several experiments on such systems, we provide the
following analysis given by Buttiker.

But before we go to Buttikers analysis, it is worth mentioning that Eq. 27 does tell us something very important.
From Eq. 27 we can write for the resistance
$$
R^{hypothetical}_{four-probe}=\frac{h}{e_0^2}\frac{1-|t(E)|^2}{|t(E)|^2}
$$
$$
R^{hypothetical}_{four-probe}=\frac{h}{e_0^2}\frac{1}{|t(E)|^2}-\frac{h}{e_0^2}
$$
Now we can see from Eq. 14 that for a perfect conductor of length $l$, there is no reflection and so $|t(E)|^2=1$, implying
there is a contact resistance of
$\frac{h}{e_0^2}$ where the classical source reservoir meets the lead.
So to this hypothetical sample resistance if we add this contact resistance then we get back Eq. 14.
$$
R^{hypothetical}_{four-probe}+\mbox{real contact resistance}=\frac{h}{e_0^2}\frac{1}{|t(E)|^2}-\frac{h}{e_0^2}+ \frac{h}{e_0^2}
$$
$$
R^{hypothetical}_{four-probe}+\mbox{real contact resistance}=\frac{h}{e_0^2}\frac{1}{|t(E)|^2}=\mbox{real two probe resistance of Eq. 14}
$$
The contact resistance and the sample resistance are in series and just add up like resistors in series.
The contact resistance can always be measured by measuring the resistance of a perfect conductor but the sample resistance
$R^{hypothetical}_{four-probe}$ cannot be measured separately as we will show below, but can be only thus inferred from Eqs. 14 and 27.

Measuring the two probe resistance and the contact resistance always involves two
probes with two contacts.
Thus we need at least two contacts to talk of a resistance,
and hence there is no way to know if the contact resistance is equally shared by the two contacts or it appears only at the source contact
or only at the drain contact.
Ref. \cite{sto} 
argue that the drain contact is reflectionless
and hence resistance less. Their idea is that drain contact resistance can be thought of as a
current carrying narrow wire
with well spaced levels injecting the current into a wide wire with close by levels. This suggests the idea that when the close by countable
levels come very close then they form a continuum of levels.
It is mathematically wrong to say that
countable states can come close together to form the continuum. Besides, continuum of states do not necessarily
make classicality emerge
and it is an unsolved problem in physics as to how classicality emerges in the reservoirs. It will be theoretically impossible to
show or argue that the wide conductor can gradually become a classical reservoir.
An experiment can only measure the two probe conductance of a sample that consist of a narrow conductor
and a wide conductor joined together between two contacts, even though the narrow-wide junction could be
fully transmitting.
These scientific explorations come from the idea that the universe is fundamentally
quantum in nature and classicality emerges.
For a comprehensive understanding of the experimentally proven
Landauer-Buttiker formalism we have to discard this view.

Noting the fact that the contact resistance is in terms of fundamental constants that can be measured to an accuracy
of one part in a billion we can not assume mesoscopic phenomenon is an approximation to condensed matter phenomenon.
In the next sub-section, we will be specifically discussing a regime in which WKB approximation fails miserably
and we will show that in that regime, there is a new origin to
local states compatible with Lorentz's symmetry and hence subject to Pauli exclusion principle.
This will therefore make it obvious that classical laws can co-exist with quantum laws without
contradicting each other and without having one to evolve into the other.
This will also solve the quantum measurement problem.
Let us see below what will be measured if we try a four probe measurement.

\textbf{\textit {Multi probe conductance:}} We put voltage and current probes in the same footing. Which means we consider a mesoscopic sample
connected to multiple leads and we have not yet fixed which lead will be used as a current probe and which as a voltage probe. They will
all cause decoherence (phase randomization) in the same manner, that is in terms of causing decoherence, all leads are equivalent.
Devices like voltmeter or ammeter are classical devices, reading off the
system after the necessary decoherence has happened. There will be no need for collapse as we will see.
Collapse of wave-function will not be random and hence
not a consequence of the decoherence but rather a consequence of the fact that a quantum
event can only be defined between two classical events in space-time. Classical events in space-time
are all predetermined according to theory of relativity. Hence
let us say that there are multiple leads attached to a mesoscopic sample that are labeled $p$, $q$, etc.
Let us focus on the lead $p$ which is attached to a
reservoir that has a chemical potential $\mu_p$. This reservoir injects at a rate of $ne_0v$
number of electrons into this lead at a point in the lead attached to it, that is determined by the LPDOS with Pauli
exclusion principle applied to the local partial state at that point at an energy $E$.
Thus the injected current into the mesoscopic sample is determined by the injecting lead alone, which was assumed
by Landauer without any reason, and one can define an injectance of this lead with respect to the full system
to which it is attached with several outgoing leads, wherein the injectance is like a resistance to this
injected current \cite{kan}.
Thereby, we write
$$
(dn)e_0v=\frac{dn}{dE}e_0vdE=
\rho_{lpd}^{pl}(+k)e_0vdE=
\frac{1}{hv}e_0vdE=\frac{e_0}{h}dE=dJ
$$
Here,
$\rho_{lpd}^{pl}(+k)$ is the LPDOS for the $+k$ mode for electrons incident from source reservoir at a point in the $p$th lead.
Since the two probe formula agrees with experiments, we start our arguments by removing all other leads except $p$ and $q$.
Current flowing out from lead $p$ to lead $q$ will be
$$
J^u_{qp}= \frac{ e_0}{h} |t_{qp}|^2(\mu_p - 0)
$$
Hence we are considering the entire energy window
$(\mu_p - 0)$, that is
up to some reference point which is taken as zero.
Now $(\mu_p - 0)$ can be a larger energy window which justifies $dJ\rightarrow J$, but we will track only the linear term that
matters in Eq. 20.
Thus for the two probe case,
when we calculate net current flowing from lead $p$ to lead $q$, (the contact resistance will cancel below $\mu_q$
but that does not mean that the contact resistance is not shared between the two contacts)
$$
J_{qp}=J^u_{qp}-J^u_{pq}=\frac{ e_0}{h} |t_{qp}|^2(\mu_p - 0)-\frac{ e_0}{h} |t_{pq}|^2(\mu_q - 0)
$$
which is a characteristic current response for the system due to $\mu_p$ and $\mu_q$ alone with other probes removed.
It is to be noted that $|t_{qp}|^2(\mu_p - 0)$ is a notation for
$|t_{qp}|^2 \Delta \mu_p$ and that
$|t_{qp}|^2(\mu_p - 0)\neq |t_{qp}|^2(\mu_p)- |t_{qp}|^2(0)$.
Never the less the
reference chemical potentials will not contribute because
we expect that for $\mu_p=\mu_q\neq 0$ we should get $J_{qp}=0$ as a consequence of detailed balance,
and hence,
\begin{equation}
|t_{pq}|^2=|t_{qp}|^2\neq 0
\end{equation}
Since this is true for all
$\mu_p=\mu_q$, Eq. 28 will be generally true. Meaning Eq. 28 will hold even for
$\mu_p\neq\mu_q$ because internal details of the system determine scattering cross section and not by imposed external condition of
$\mu_p=\mu_q$. Besides Eq. 28 should be
independent of the direction of magnetic field because for any magnetic field, for
$\mu_p=\mu_q$, we should get
$J_{qp}=0$
(this again only implies that contact resistance cancels for $\mu_p=\mu_q$, but does not
imply that it is not shared
between the two contacts).
For $\mu_p=\mu_q$
there will be no current in the system even if no measurements are made.
Thus we recover the two probe formula
\begin{equation}
J_{qp}= \frac{ e_0}{h} (|t_{qp}|^2\mu_p - |t_{pq}|^2 \mu_q)
= \frac{ e_0}{h} |t_{qp}|^2(\mu_p - \mu_q)
\end{equation}
Now when we have more than two leads, we will say that Eq. 29 is valid with two physical conditions imposed.
First is that, when there are many leads, we do not commit that $\mu_p > \mu_q$ because the opposite is also possible,
in other words the sign of $J_{qp}$ is arbitrary. Second is that, between any two $\mu_p$ and $\mu_q$
there is no detailed balance. However, a detailed balance will act among all the leads collectively,
which we do not know in details. We will expect that these two matter of details will be taken care of
by imposing a fact that is always found in experiments. That is for more than two leads
\begin{equation}
|t_{pq}(+B)|^2=|t_{qp}(-B)|^2\neq 0
\end{equation}
meaning transmission
amplitude from $p$ to $q$ is the same as the transmission amplitude from $q$ to $p$ with the
direction of magnetic field reversed.
Ref \cite{dat} give some arguments as to why this can happen (see page 123) but why the two probe situation and
multi-probe situation is different can be only understood by looking at Feynman paths and their orientation with respect to
the magnetic field and position of leads. For example, if there are only three leads, then for a fixed direction of
current, if we reverse the magnetic field, then the position of the third lead also has to be adjusted to get
combination of identical Feynman path lengths.

We will see that we will get new kinds of reciprocity
relations when there are many leads, that are different from Onsager reciprocity relation that was derived from the principle of
detailed balance in the thermodynamic limit.
Onsager had shown that for resistivity, $\rho_{xx}(B)=\rho_{xx}(-B)$ and that is what lead researchers to believe that
resistance will remain the same
on reversing the magnetic field, specially because that is also what was found from the two
probe formula. But
this was not found for multi-lead mesoscopic systems and strengthens the fact
that for mesoscopic systems we cannot define a resistivity, as stated in the introduction.
For only two leads, since the measured resistance is the sample resistance plus the contact resistance,
we always get
$|t_{pq}(B)|=|t_{pq}(-B)|$
as expected on physical grounds, because
reversing the magnetic field is just looking at the sample from above or bottom.
In general
$|t_{pq}(B)|\neq |t_{pq}(-B)|$ in presence of more leads than just $p$ and $q$.

In summary, for more than two probes, we will start from Eq. 29 wherein the direction of $J_{qp}$
is arbitrary and there is no detailed balance between any two $p$ and $q$.
If we can write the electrostatic potential in the $p$th reservoir as
\begin{equation}
U_p=\frac{\mu_p}{e_0}
\end{equation}
then
$$J_{qp}= \frac{ e_0^2}{h} (|t_{qp}|^2U_p - |t_{pq}|^2 U_q)
$$
\begin{equation}
J_p=\sum_qJ_{qp}=\sum_q(G_{qp} U_p - G_{pq}  U_q)
\end{equation}
and this is measurable, where
\begin{equation}
G_{pq}=\frac{e^2_0}{h} |t_{pq}|^2
\end{equation}
If we make $U_q=U_p$ for all $q$ in Eq. 32, then we must get $J_p=0$ for
all magnetic fields, which gives the following sum rule (which will be explicitly
shown below as we discuss three probe conductance).
$$
0=\sum_q(G_{qp} U_p - G_{pq}  U_p)
$$
\begin{equation}
\mbox{or}\;\;\sum_q G_{qp} =\sum_q G_{pq}
\end{equation}
This is true for all $p$ and $U_p$ and hence is an identity.
Eqs. 30 and 34 taken together can be considered new reciprocity relation.
That is
$$
\sum_q G_{qp}(+B) =\sum_q G_{pq}(+B)
$$
$$
G_{qp}(+B) =G_{pq}(-B)
$$
This is never found to be violated in an experiment.
Substituting Eq. 34 into Eq. 32 we get
\begin{equation}
J_p= \sum_q G_{pq} (U_p - U_q)
\end{equation}
Now if the $p$th lead needs to be used as a voltage probe then for $U_p \neq U_q$, it should not draw any current or $J_p$=0.
This condition yields a solution to Eq. 35 giving the following $U_p$ which is what would
be the voltage at the point where the $p$th lead attaches to the sample.
\begin{equation}
U_p =\frac{\sum_{q \neq p} G_{pq} U_q}{\sum_{q \neq p} G_{pq}}
\end{equation}
This is true for arbitrary number of $p$ and $q$ and so can be applied to a 3 probe set up or a 4 probe set up, but does
not apply to the two probe case.

Let us talk of a three probe conductance instead of a four probe conductance because that is very informative.
This means $p$ and $q$ take values from the set $\{1,2,3\}$.
Which means lead 1 is from the source reservoir and lead 3 is going into the sink reservoir. That also means $U_1$ and $U_3$
are applied by an experimentalist while he is trying to measure $U_2$ which gives the voltage of the floating point to which
lead 2 attaches to the sample.
In Eq. 35 let us put $p=1$ and let there be three leads only, which gives.
$$
J_1=G_{11}(U_1-U_1) + G_{12}(U_1-U_2) +G_{13}(U_1-U_3)
$$
Or
$$
J_1=(G_{12}+G_{13})U_1- G_{12}U_2 - G_{13} U_3
$$
Similarly
$$
J_2=-G_{21}U_1+(G_{21}+ G_{23})U_2 - G_{23} U_3
$$
Similarly
$$
J_3=-G_{31}U_1-G_{32}U_2+ (G_{31}+ G_{32}) U_3
$$
Therefore
\begin{equation}
    \begin{bmatrix}
        J_1 \\
        J_2 \\
        J_3
    \end{bmatrix}
=
\begin{bmatrix}
    G_{12} + G_{13} & -G_{12} & -G_{13}\\
    -G_{21} & G_{21}+G_{23} & -G_{23} \\
    -G_{31} & -G_{32} & G_{31} +G_{32}
\end{bmatrix}
\begin{bmatrix}
    U_1\\
    U_2\\
    U_3
\end{bmatrix}
\end{equation}
To cast this in the form of a matrix equation, we define
$$
G_{12} + G_{13} = -G_{11}
$$
$$
G_{21} + G_{23} = -G_{22}
$$
$$
G_{31} + G_{32} = -G_{33}
$$
Hence
\begin{equation}
    \begin{bmatrix}
        J_1 \\
        J_2 \\
        J_3
    \end{bmatrix}
=
\begin{bmatrix}
    -G_{11} & -G_{12} & -G_{13}\\
    -G_{21} & -G_{22} & -G_{23} \\
    -G_{31} & -G_{32} & -G_{33}
\end{bmatrix}
\begin{bmatrix}
    U_1\\
    U_2\\
    U_3
\end{bmatrix}
\end{equation}
which gives
\begin{equation}
J_1=-G_{11}U_1 -G_{12}U_2 -G_{13}U_3 
\end{equation}
Physically we may expect that if all voltages are the same then we will get no currents.
That is if we set $U_1=U_2=U_3\neq 0$ then $J_1=J_2=J_3=0$ must happen.
That implies, from Eq. 39
\begin{equation}
0=-G_{11}U_1 -G_{12}U_1 -G_{13}U_1
\end{equation}
Hence, from Eq. 38
\begin{equation}
G_{11}+G_{12}+G_{13}=0 \;\;\mbox{and}\;\; -G_{11}= G_{12} +G_{13}
\end{equation}
\begin{equation}
G_{21}+G_{22}+G_{23}=0 \;\;\mbox{and}\;\; -G_{22}= G_{21} +G_{23}
\end{equation}
\begin{equation}
G_{31}+G_{32}+G_{33}=0 \;\;\mbox{and}\;\; -G_{33}= G_{31} +G_{32}
\end{equation}
which justifies the definition of $G_{11}$ and also partly explains Eq. 34.
Next we expect Kirchoff's law to hold, which means $J_1+J_2+J_3=0$ for arbitrary $U_i$s.
\begin{equation}
J_1+J_2+J_3=0
\end{equation}
So from Eq. 38 we get
\begin{equation}
-G_{11}U_1 -G_{12}U_2 -G_{13}U_3 
-G_{21}U_1 -G_{22}U_2 -G_{23}U_3 
-G_{31}U_1 -G_{32}U_2 -G_{33}U_3 = 0
\end{equation}
\begin{equation}
-(G_{11}+G_{21}+G_{31})U_1
-(G_{12}+G_{22}+G_{32})U_2
-(G_{13}+G_{23}+G_{33})U_3 =0
\end{equation}
Since this is true for any arbitrary $U_i$s
\begin{equation}
G_{11}+G_{21}+G_{31}=0
\end{equation}
\begin{equation}
G_{12}+G_{22}+G_{32}=0
\end{equation}
\begin{equation}
G_{13}+G_{23}+G_{33}=0
\end{equation}
This justifies the rest of Eq. 34.
Note that the $G$s are determined by the system. The voltages and currents are determined by the $G$s.
Some of the voltages may be applied and the others are response determined by the response functions $G$s.
Now we have decided for the system in Fig. 2,
which is a current lead and which is a voltage lead. Lead 2 is the voltage lead which
will measure the voltage of the floating point to which it is attached and $U_2$ will be adjusted such that
$J_2$ becomes zero and this adjusted value will give the voltage of the floating point to which lead 2 is attached.
While fixed $U_1$ and $U_3$ is applied to
provide a bias that drives a current from 1 to 3, the basic property of a voltage probe is that it should not
draw any current.
One can choose to set $U_3=0$ without any loss of generality and write from Eq. 36.
\begin{equation}
U_2=\frac{G_{21}}{G_{21}+ G_{23}}U_1
\end{equation}
From Eq. 37 we can write for $U_3=0$
\begin{equation}
J_2=-G_{21}U_1+(G_{21}+ G_{23})U_2
\end{equation}
\begin{equation}
J_3=-G_{31}U_1-G_{32}U_2
\end{equation}
We need not consider $J_1$ because by knowing $J_2$ and $J_3$, we can know $J_1$ from Eq. 44.
Since the voltage probe does not draw any current, $J_2=0$ and so Eq. 51 gives back Eq. 50.
From Eq. 52 for $J_3=J_1$ we get
\begin{equation}
\frac{J_1}{U_1}=-G_{31}-G_{32}\frac{G_{21}}{G_{21}+ G_{23}}
\end{equation}
This is always arbitrary to a sign depending on convention that current entering the system is positive and current leaving the system is
negative or vice versa. In fact Eq. 44 may be written as $J_1+J_2-J_3=0$.
Or we can also express it as
\begin{equation}
\frac{J_1}{U_2}=
\frac{G_{21}+ G_{23}}{G_{21}}\bigg(-G_{31}-G_{32}\frac{G_{21}}{G_{21}+ G_{23}}\bigg)
\end{equation}
Since Eqs. 53 and 54 has been experimentally verified time and again, in presence or absence of magnetic fields, one can say that
Pauli exclusion principle associated with LPDOS has also been verified experimentally.

\subsection{Demonstration of $\rho_{lpd}$ from $G_{\alpha \beta}$s}

For the purpose of this demonstration we consider the three probe (three leads attached to a mesoscopic sample)
set up as described below with respect to Fig. 2, but the logic can be extended to arbitrary number of leads.
The mesoscopic sample is the shaded region which is typically made of
semi-conductor or metal. The leads $\gamma$
or $\alpha$ are current leads that may be made up of the same material.
Lead $\beta$ is an STM tip
that can attach to different points in the sample and
is such that we can vary its proximity to the sample.
The STM tip can be set up in many different ways but we will focus on the particular case considered by
Buttiker and his co-workers \cite{gra}. It will be clear as to why they could only understand the semi-classical regime
of that set up.
That is the set up when the STM tip is weakly coupled to the sample at a point $\textbf r$ through a tunneling barrier.
While Eqs. 53 and 54 is measured by making $J_2=0$ by adjusting $\mu_2$, one can also earth the STM tip as shown in Fig. 3 and measure
the coherent current from reservoir 1 to reservoir 2. This coherent current can demonstrate $\rho_{lpd}$ as will be shown in this subsection.
This coherent current can be again measured by measuring scattering probabilities as is evident from Eq. 13.

In axiomatic quantum mechanics, any local state will be disturbed by measurements
in an unpredictable manner and the theory of quantum mechanics rests on the fact that there can only be a random outcome. There is no
theory for this measurement related randomness in quantum mechanics, as opposed to the dynamics prior to measurements
for which probabilities can be calculated.
A good example to see this are what is known as snapshots of chaotic billiards.
Once again Buttikker and
his coworkers \cite{gra} could demonstrate a connection between measured currents and LPDOS in the semi-classical regime
for the typical set up of Fig. 2. The real
issue is in the quantum
regime and that remained puzzling as LPDOS could become negative and they could not make sense of it. We had shown that negative
LPDOS is consistent with negative signal propagation time or time travel \cite{kan}. In the previous subsection
we also showed that LPDOS can answer the unanswered questions related to why Landauer-Buttiker
formalism works so well. In this sub-section, we will demonstrate that
the relation between LPDOS and measured scattering probability, as derived by Buttikker and co-workers make perfect sense even when we get
negative LPDOS. 
It will be thus clear that the possibility of a deterministic local state with a
LPDOS also solve the quantum measurement problem as will be demonstrated in this sub-section.
Just as in principle, one can solve Newtons equation for $10^{23}$ particles to derive classical statistical mechanics,
one can in principle calculate LPDOS for a mesoscopic sample with $10^{23}$ leads attached, to deterministically predict
every single electron scattering outcome.
As to why the uncertainty principle will not affect this deterministic outcome is discussed in \cite{kan} (see Eqs.
3.56 and 3.57 therein with the discussion below it), but the intuitive answer is that local states do not
appear in axiomatic quantum mechanics but the axioms do not rule out its possibility by Goedel's incompleteness
theorem.
This applies to any other theorem that can be proved within the axiomatic framework of quantum mechanics, like for
example the Bell's theorems.

\begin{figure}[bt]
\centering
\includegraphics[width=1.2\textwidth, keepaspectratio]{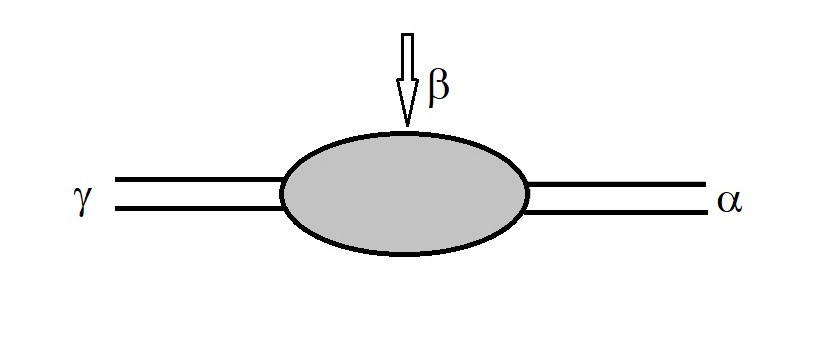}
\captionsetup{labelformat=empty}
\caption{\label{fig2}
Fig. 2. The basic set up for a three probe Landauer conductance where there are
only two fixed leads indexed $\gamma$ and $\alpha$ apart from the STM tip $\beta$.}
\end{figure}

\begin{figure}[bt]
\centering
\includegraphics[width=0.8\textwidth, keepaspectratio]{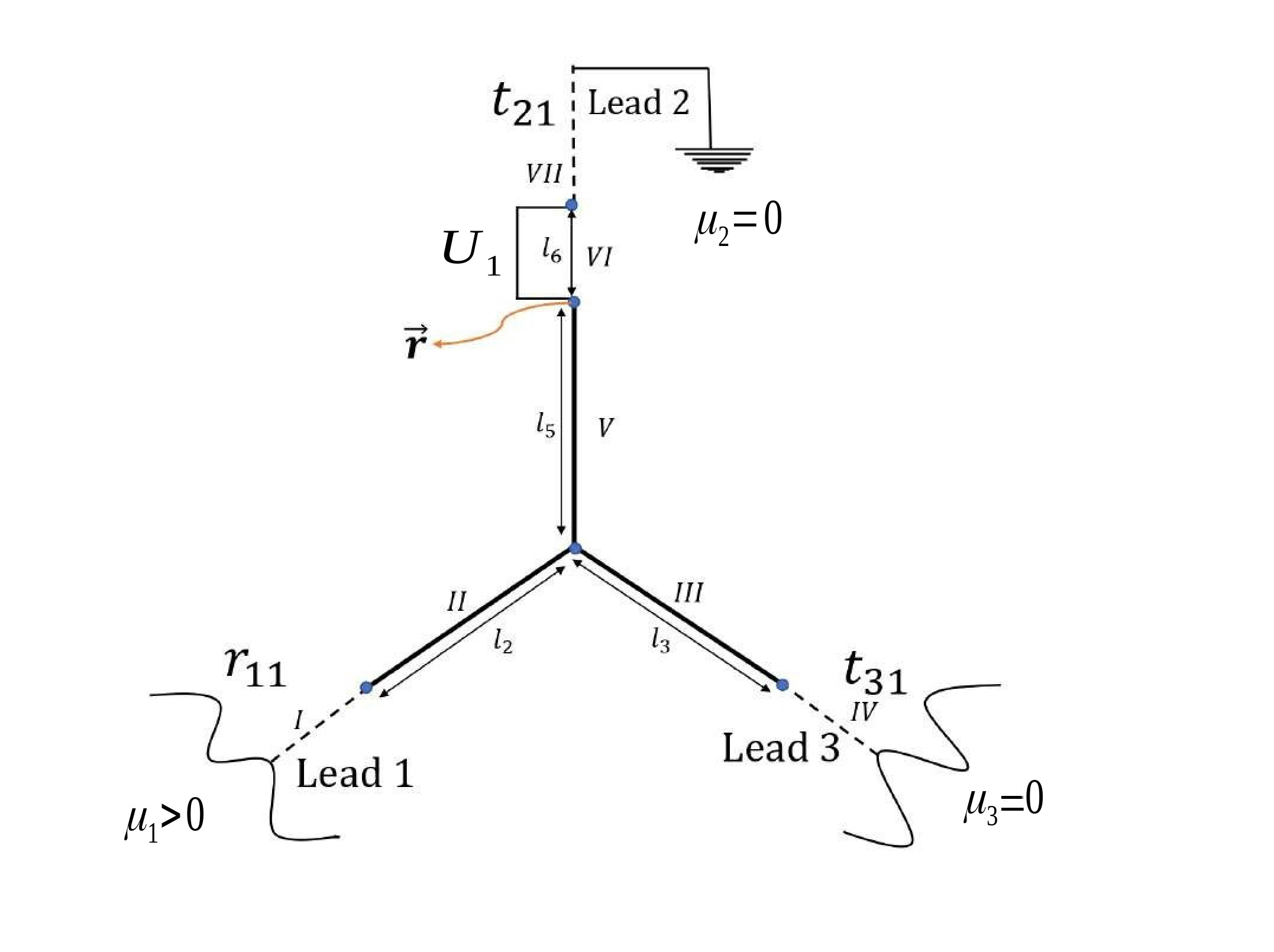}
\captionsetup{labelformat=empty}
\caption{\label{fig3}
Fig. 3. The sample is the three prong potential shown by the solid lines and the entire system consist of the sample
connected to three reservoirs via three leads.
Different regions of the system is marked by Roman numbers, I, IV and VII being the leads, shown by dashed lines.
Lead 2 is similar to the lead $\beta$ in Fig. 2 but earthed and the chemical potential of lead 3 is also set to
zero. In that case, conventionally, a small
positive chemical potential at lead 1 will inject a current
while the two other leads
carry current away from the sample.
Lead 2 connects to the sample through a tunneling barrier shown as region VI.
Lengths of different regions is shown as $l_2$, $l_3$ etc. This system is a 1D version of the system in Fig. 2 where
lead $\gamma$ is renamed as lead 1, etc, and as a result $t_{31}\equiv s_{31}$,
$t_{21}\equiv s_{21}$, and
$r_{11}\equiv s_{11}$.}
\end{figure}

Let chemical potential of left reservoir connected to lead $\gamma$ be $\mu_\gamma>0$,
chemical potential of right reservoir connected to lead $\alpha$ is $\mu_\alpha=0$,
and the other end of the the lead $\beta$ is earthed. Essentially, $\mu_\gamma$ and $\mu_\alpha$
are fixed for all incident energies (or other parameters),
but $\mu_\beta$ can floatingly adjust to zero as the potential at the point $\textbf r$ where
the STM tip touches the sample depend on many parameters. For incident energy $E$ such that
$0=\mu_\alpha<E<\mu_\gamma$, only lead $\gamma$ injects a current according to our classical notions.
For a mesoscopic sample, according to Landauer-Buttiker formalism, this defines
a quantum mechanical scattering problem.
Such a set up can be obtained in the laboratory and
we want to address
the coherent current flowing from $\gamma$ to $\alpha$. One can derive \cite{lar} that in such a situation
\begin{equation}
|s_{\alpha \gamma}'|^2- |s_{\alpha \gamma}|^2 = - 4\pi^2 \mid{t_\beta}\mid^2 \nu_\beta \rho_{lpd}(E, \alpha, \textbf{r}, \gamma)
\end{equation}
where
\begin{equation}
\rho_{lpd}(E, \alpha, \textbf r, \gamma) = -\frac{1}{2 \pi e_0} \; |s'_{\alpha \gamma}|^2 \; \frac{\delta \theta_{s'_{\alpha \gamma}}}{\delta U({\bf{r}})}
\end{equation}
To make a connection with Eq. 1, it is suitable to define
$$
\rho_{lpd}^r=\rho_{lpd}(E, 1, \textbf r, 1) \;\;\mbox{and}\;\;
\rho_{lpd}^t=\rho_{lpd}(E, 3, \textbf r, 1)
$$
which are local partial density of states (LPDOS) for this system, where
$U(\textbf r)$ is the electrostatic potential at the point $\textbf r$
and $e_0$ is the charge of the particles.
\begin{equation}
\theta_{s'_{\alpha \gamma}}=ArcTan\bigg[\frac{Im[s'_{\alpha \gamma}]}{Re[s'_{\alpha \gamma}]}\bigg]
\end{equation}
and
$\frac{\delta \theta_{s'_{\alpha \gamma}}}{\delta U({\bf{r}})}$ is a functional derivative.
The parameter $t_\beta$ controls coupling of states of the sample with the states at the STM tip for which $\nu_\beta$ is the relevant density of states (DOS).
For a commercially available STM tip $t_\beta$ and $\nu_\beta$ are just
parameters, either specified by the manufacturer or to be ingeniously approximated by the user.
Thus, $s_{\alpha \gamma}'$ is the scattering matrix element for scattering from $\gamma$ to $\alpha$ when the STM tip is drawing a current given by the
RHS of Eq. 55.
When the STM tip is removed by $t_\beta\rightarrow 0$ then this scattering amplitude will be
$s'_{\alpha \gamma}=s_{\alpha \gamma}$.
Coherent currents are given by $\frac{e_0}{h}|s'_{\alpha \gamma}|^2$ and
$\frac{e_0}{h}|s_{\alpha \gamma}|^2$ in the two situations.

In earlier works, Eq. 55 was studied in the semi-classical regime wherein $\rho_{lpd}$ is
positive definite, which limits its value and applicability in the quantum regime.
The quantum regime remained unclear specially because in the quantum regime $\rho_{lpd}$, as calculated
from Eq. 56, can become negative.
We will show that Eq. 55 remains valid even when
$\rho_{lpd}$ becomes negative in a quantum regime.
For this it is extremely important to apply wave-mechanics or von Neumann quantum mechanics for the region between the two reservoirs,
that is the mesoscopic system.
The key idea needed to show this is to realize the fact that in the quantum regime of 1D, 2D
and 3D, $\rho_{lpd}$ cannot really be negative although sometimes it misleads one to think so.
Real negativity of $\rho_{lpd}$ is a specialty of the mesoscopic regime when we encounter Fano resonances \cite{kan}.
This is the primary reason why earlier works could not understand the quantum regime and to be demonstrated below.

Intuitively speaking, $\rho_{lpd}$ is positive and the LHS of Eq. 55 is negative accounting for loss of coherent electrons to the earthed lead $\beta$.
Therefore, any change of sign in $\rho_{lpd}$ will counter-intuitively result in a gain of coherent current due to the intervention
of the earthed STM tip on a quantum state.
That can be regarded as a supporting evidence for time travel (like a quantum
state getting younger with respect to coordinate time) because any intervention on a quantum state at an intermediate point,
by an STM tip is supposed to only decohere the state.
We will clarify as to why and when such counter-intuitive effects can be seen in real systems.
Since LHS of Eq. 55 can be defined in quantum mechanics and can also be
measured in the laboratory and so if the equality in Eq. 55 can be established then one can affirm that RHS
is an objective reality that can be found in nature, although we cannot make sense of it within the axiomatic framework
of quantum mechanics.

In this paragraph we discuss that $\rho_{lpd}$ as a local object solves the problem of collapse of wave-function by making it redundant,
with measurements defined deterministically in the classical world of the reservoirs, while the time evolution
of observables in the region between the reservoirs is determined by quantum mechanical scattering.
At a point $\textbf r$ in the shaded region,
there will be electrons in a linear superposition of states.
One may say that all these electrons came from lead $\gamma$ (assume for the time being that $\rho_{lpd}$ is positive definite)
but some may be going to lead $\alpha$ (we call them dancing cats), some may be going to lead $\beta$ (we call them sleeping
cats) and some may be getting reflected back to lead $\gamma$ (we call them beaten cats). At the point $\textbf r$,
we have a linear superposition of dancing, sleeping and beaten cats. In formal quantum mechanics, we cannot look
at the cats at the point $\textbf r$ and say which one is dancing (meaning going to lead $\alpha$) etc.
But we can now say something about only the dancing cats. They spend a time
$-\frac{h}{2 \pi} \; \frac{\delta \theta_{s'_{\alpha \gamma}}}{\delta U({\bf{r}})}$, at the point $\textbf r$.
At zero temperature there are $|s'_{\alpha \gamma}|^2$ of these dancing cats at that point and we can average over
only the dancing cats to define a local partial density of states
given by
$\rho_{lpd}(E, \alpha, \textbf r, \gamma) = -\frac{1}{2 \pi} \; |s'_{\alpha \gamma}|^2 \; \frac{\delta \theta_{s'_{\alpha \gamma}}}{\delta U({\bf{r}})}$.
That means we can selectively average over the dancing cats at the point $\textbf r$ and determine physical effects due to
the dancing cats alone. Like for example the LHS of Eq. 55 is physical and is determined by
the dancing cats, only thing that remains to be verified is that the equality in Eq. 55 holds.
In formal quantum mechanics we cannot do this selective averaging for the dancing cats alone that
are in a linear superposition with the sleeping and beaten cats.
Note that
$|s'_{\alpha \gamma}|^2$ and
$\theta_{s'_{\alpha \gamma}}$ are measured at the junction of the sink reservoir which is a classical system.
Probability is well defined for classical systems and measured
$|s'_{\alpha \gamma}|^2$ is a probability wherein measurements do not care how the probability is defined theoretically. 
Similarly, according to the notion of physical clock,
$\theta_{s'_{\alpha \gamma}}$ is defined as angular displacement of a precessing dipole.
A perfectly logical connection can be made between quantum mechanical scattering phase shift
$\theta_{s'_{\alpha \gamma}}$ and well defined angular displacement of a precessing dipole \cite{kan}.
Thus the measurables are not random outcomes induced by collapse of wave-function by the intervening STM tip at
$\textbf r$, but can be calculated precisely
in the Landauer-Buttiker formalism wherein the punchline is that a quantum event can only be defined between two classical events
in the reservoirs that perfectly help define a local time and a local state at the point $\textbf r$.
We know that von Neumann quantum mechanics has one mathematical framework for scattering states and bound states.
For any scattering phenomenon the LPDOS can be defined and calculated
as bound states can be seen as a special case of the scattering states.
To make the connection between scattering matrix and LPDOS, we need the Landauer-Buttiker formalism and a physical clock.
So we believe that this solves the measurement problem in general and we have to see how the Landauer-Buttiker
formalism can be extended to other practical cases.
At various places in this work,
we have sketched out this road map as to what is a general mesoscopic system and how any other system can be seen as a modification
of it. That is why we have discussed von Neumann formalism, Dirac's approach, Heisenberg's approach, S. Datta's approach,
reservoirs being in the domain of relativity, etc.

Thus we intend to justify Eq. 55 for the quantum regime, specially the situation when $\rho_{lpd}$ becomes negative.
Ref. \cite{gra} gives a derivation of Eq. 55, we will give a physical topological argument here which will make it
clear why and when we can generally expect the equality in Eq. 55 in spite of $\rho_{lpd}$
being negative. That will be followed by a numerical demonstration.
Note that there are two DOS that feature in Eq. 55, that are $\nu_\beta$ and an $\textbf r$ dependent DOS that is $\rho_{lpd}$.
There is a factor $\frac{1}{2\pi}$ for each of them that cancels the factor $4 \pi^2$.
This is the $\frac{1}{2\pi}$ factor that appears in Eq. 56.
There can be a very simple interpretation for this factor as follows.
LHS of Eq. 56 is a DOS that can accommodate electrons that are countable
but $2 \pi \rho_{lpd}$ is uncountable, the RHS of Eq. 56 being
in terms of $\theta_{s_{\alpha \gamma}}(E)$ corresponding to a continuous rotation.
This also make it essential that if only one half of a Riemann surface is involved then rotation angle can only
be maximum $\pi$ instead of
$2\pi$. So winding number in Eq. 56 as a conserved quantity
in the two cases will differ by a factor of 2. On the other hand, scattering probabilities
$|s'_{\alpha \gamma}|^2$ and $|s_{\alpha \gamma}|^2$ are just numbers,
are well defined in quantum mechanics, and when
multiplied by a factor $\frac{e_0}{h}$ they give the measured coherent current from $\gamma$ to $\alpha$.

Let us consider the system shown in Fig. 3. First of all it is a system for which exact quantum mechanical calculations can be made
and it is a system that exhibit pronounced Fano resonances.
The system in Fig. 2 will generally show such resonances was shown in earlier works \cite{kan}.
In this sense the system in Fig. 2 at Fano resonances with the modification
that the upper lead is weakly coupled to earth via a strong potential can be represented by the 1D system in Fig. 3.
Besides, since the logic, to be presented below, is based on topology of the Argand diagram (AD) at a resonance
and the complex plane, there are only two kinds of behavior,
one being typically like that of Breit-Wigner resonance that involve the full Riemann surface, and the
other being typically like that of Fano resonance that involve one half of the Riemann surface.
Also since it is a question of whether LPDOS is present at all in a quantum system and more importantly whether its negativity is allowed in
quantum mechanics, then demonstrations in 1D quantum mechanics should suffice.
Let us drop the factor
$|t_\beta|^2 \nu_\beta$ from Eq. 55 because such factors will be only needed if we are using a commercially available STM tip,
where this factor characterizes the STM tip.
This require us to drop a $2\pi$ factor in Eq. 55 for reasons mentioned in the previous paragraph.
Which means for a first principle calculation for the system in Fig. 3
we expect (if the equality in Eq. 55 is correct) that
\begin{equation}
|s_{\alpha \gamma}'|^2- |s_{\alpha \gamma}|^2 = - 2 \pi \rho_{lpd}(E, \alpha, \textbf{r}, \gamma)
\end{equation}
Now to make it correspond to the notations in Fig. 3
\begin{equation}
|t_{31}'|^2- |t_{31}|^2 = - 2 \pi \rho_{lpd}(E, 3, \textbf{r}, 1)
\end{equation}

\begin{figure}[bt]
\centering
\includegraphics[width=1.2\textwidth, keepaspectratio]{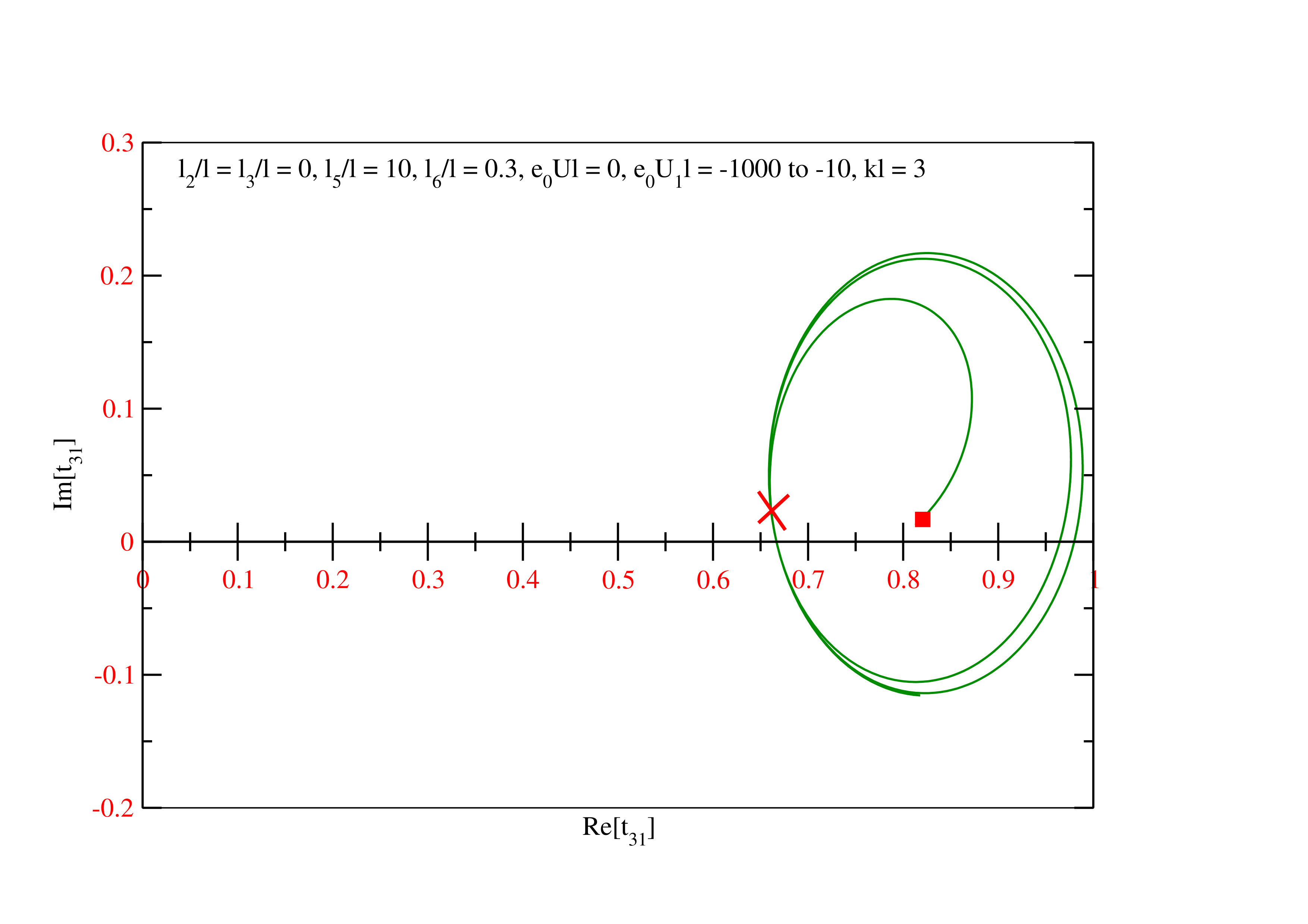}
\captionsetup{labelformat=empty}
\caption{\label{fig33}
Fig. 4. In this figure we plot the AD for $t_{31}$ of the system shown in Fig. 3
for the coupling potential $U_1$ varying in a range that give three sub-loops, all within
one Riemann surface. The starting point is marked by a small square block corresponds to $U_1l=-10000$. The end point is unmarked and
corresponds to a value $U_1l=-10$. All the sub-loops smoothly come back to a point marked by a cross. Other parameters are mentioned inside
the figure.}
\end{figure}

In Fig. 4 we plot the Argand diagram (AD) for $t_{31}$ for the system shown in Fig. 3 by varying the tunneling
potential $U_1$ which connects the three prong
potential with lead 2 that is earthed. As $U_1$ is
varied from $e_0U_1l=-10$ to
$e_0U_1l=-1000$ we plot $Im[t_{31}]$ versus
$Re[t_{31}]$, which is known as AD.
The AD diagram make smooth sub-loops within one Riemann surface and each sub-loop is due to
a Fano resonance, because they do not go around the origin
(0,0). Any physical quantity that depend on $t_{31}$ will go through a cycle over one particular sub-loop.
For one closed sub-loop generated by monotonously varying a parameter, say $U_1(\textbf r)$, the local potential at $\textbf r$.
\begin{eqnarray}
\oint_c \Delta\theta_{t_{31}} =
\oint_c \frac{\delta \theta_{t_{31}}}{\delta U_1({\bf{r}})} \; \Delta U_1({\bf{r}})=
0 \;\; \mbox{where}\;\;
\theta_{t_{31}} =  Arctan \frac{Im[t_{31}]}{Re[t_{31}]}
\end{eqnarray}
Because the AD was generated by varying $U_1(\textbf r)$, the integration measure is
$\Delta U_1(\textbf r)$ and the integration for a closed loop will be zero simply because the AD does not go around the origin.
Or
\begin{eqnarray}
\oint_c {\frac{1}{2 \pi} |t_{31}|^2 \; \frac{\delta \theta_{t_{31}}}{\delta U_1({\bf{r}})}} \; \Delta U_1({\bf{r}})=
-\oint_c {\rho_{lpd}(E,3,\textbf r, 1)} \; \Delta U_1({\bf{r}})=0
\end{eqnarray}
Similarly over the same sub-loop,
\begin{eqnarray}
\oint_{c} \frac{\delta|t_{31}|^2}{\delta U_1(\textbf r)} \Delta U_1(\textbf r)=
\oint_{c} {\Delta|t_{31}|^2}=
\oint_{c} {\Delta|Re[t_{31}]+Im[t_{31}]|^2}=0
\end{eqnarray}
Therefore from Eqs. 61 and 62 we get the equality
\begin{equation}
\oint_{c} {\Delta|t_{31}|^2}=
\oint_c {\frac{1}{2 \pi} |t_{31}|^2 \; \frac{\delta \theta_{t_{31}}}{\delta U_1({\bf{r}})}} \; \Delta U_1({\bf{r}})=
-\oint_c {\rho_{lpd}(E,3,\textbf r, 1)} \; \Delta U_1({\bf{r}})=0
\end{equation}
purely as a consequence of the topology of a complex plane $(Im[t_{31}], Re[t_{31}])$.
In this complex plane
$(Im[t_{31}], Re[t_{31}])$, the point (0,0) is a singular point where the phase 
$\theta_{t_{31}} =  Arctan \frac{Im[t_{31}]}{Re[t_{31}]}$ is undefined
and is well known for classical waves \cite{ber}.
If the Argand diagram trajectory would have enclosed the singularity, then the RHS of Eq. 60 being
a winding number, would have
been $2\pi$ instead of zero but the RHS of Eq. 62 would be still zero, in which case Eq. 63 will not be valid.
We get Eq. 63 from Eqs. 61 and 62 because due to Fano resonances, one gets loops in the AD within one
Riemann surface.
Whatever be the terms in the Hamiltonian or the equation of motion, it is the topology of the complex plane that determines
the outcome of the integration in Eqs. 61 and 62.
For the kind of Argand diagram in Fig. 4, that are smoothly closed within one Riemann surface,
the equality in Eq. 63 is for the integrals. We will address the value of the integrands latter
but the integrands will go through a positive-negative cycle over a loop simply because of the fact that
the integration over a loop will be zero and there is no surprise if $\rho_{lpd}$, which is the integrand
in Eq. 61, becomes negative. However, the question remains if it is physical and if it can be found in nature.
That can be verified by also considering its equality with the integrand in the LHS of Eq. 63 as demands Eq. 59.
Just to remind once again Ref. \cite{lar} could not make sense of the negativity of $\rho_{lpd}$, because they did not consider
Fano resonances.

Therefore, it is obvious from Fig. 4 and Eq. 63, that
$\Delta|t_{31}|^2$ as well as $\rho_{lpd}$
will go through a positive-negative cycle over a closed loop
as a consequence of the topology of the relevant complex plane, only if the AD closes within one side of the Riemann surface,
as in the case of Fig 4, the AD is restricted to the first and the fourth quadrants.
However, cycles are similar only means qualitative agreement, we have to show they are quantitatively equal to claim the equality
in Eq 59, or to claim that $\rho_{lpd}$ is physical, or to claim that the local states determine measurement outcomes.
If we want to compare the integrands in Eq. 63, then we need an extra factor of $2\pi$ as in Eq. 58 and 59
as derived in \cite{gra}. Thus (assuming $e_0=1$)
\begin{equation}
{\Delta|t_{31}|^2} \approx
2 \pi {\frac{1}{2\pi} |t_{31}|^2 \; \frac{\delta \theta_{t_{31}}}{\delta U_1({\bf{r}})}} \; \Delta U_1({\bf{r}})
\end{equation}
Viewing a derivative as an effect of an infinitesimal change in $U_1$ we get
\begin{equation}
|t'_{31}|^2 - |t_{31}|^2 \approx
|t'_{31}|^2 (\theta_{t'_{31}} - \theta_{t_{31}})
\end{equation}
where primed quantities and un-primed quantities are calculated for an infinitesimal difference of $U_1$.
If we
want to compare integrands then we can only write an approximate equality because
we are comparing two different
objects over one sub-loop.
There will always be in the least a phase difference between the two quantities.
But if $\rho_{lpd}$ determines the measured quantity
$|t'_{31}|^2 - |t_{31}|^2$ then the two sides of Eq. 65 should agree in amplitude.

In Fig. 5 we plot LHS (solid curve) and RHS (dashed curve) of Eq. 65 where the primed values and un-primed values are again for small differences in
$U_1$ as the incident wave vector $k$ is varied and that is the horizontal axis.
Initially the two curves are a little different but as
$k$ increases, effects of dispersion is minimized, and magnitudes of the two curves are similar with a phase difference between them.
The oscillations of both the curves are due to the smooth cyclic nature of the AD in Fig 4. In other words, the underlying principle
behind the oscillation of both curves is the same. The two curves oscillate as a result of quantum interference. But the symmetric
oscillation between positive and negative values and the similarity between the two curves can only be noted if there are
well defined Fano resonances and signify the fact that the RHS of Eqs. 58 and 59 are physical.

Note that for Fig. 5 we have set $l_2=l_3=0$ which implies that we have only Fano resonances \cite{kan}. In case of Fig. 6 we introduce
a finite value for them, that is
$l_2=l_3=l_5=1$ which means there is an interplay of Fano and Breit-Wigner type resonances.
However, the equality of Eq. 65 is still qualitatively and quantitatively valid except that
the dashed curve is slightly pushed up with respect to the solid curve, the reason for which will be
clear from the next case. Although there are only two possibilities, that are, the AD remain restricted
to one Riemann surface without enclosing the origin (as in the case of Fig. 4) or the AD goes around the origin and goes to
higher Riemann surface (as in the case of Fig. 7), there interplay can lead to diversity. For example, in 
Fig. 7 we show the AD for a set of parameters. In this case if we follow the solid curve,
the AD is initially restricted to one Riemann surface, starts from point A and
gets very close to the origin at the point B but does not enclose the origin, and then goes through the points CDEFG enclosing the origin.
At the point F it comes back to the point C making CDEF a closed loop encircling the origin but not coming back smoothly as in the case of
Fig. 4 and instead forming a cusp at C or F. For the choice of parameters
in Fig. 7, if we try to test the equality in Eq. 65 then we
show in Fig. 8 that we still have very good quantitative agreement but qualitatively
the LHS of Eq. 65 (solid curve) is very different from the RHS of Eq. 65 (dashed curve).
Such qualitative disagreement completely overshadows the quantitative agreement between the solid and dashed curves in Fig. 8 and that
can be confusing. However, if we look at the topology of the complex plane $\bigg(Im[t_{31}],Re[t_{31}]\bigg)$
then it is again possible to read off the solid curve by looking at the dashed curve. For that one has to notice that
as we move along the AD  by varying the energy (or wave-vector) going through the points CDEF then
$\theta_{t_{31}} =  Arctan \frac{Im[t_{31}]}{Re[t_{31}]}$ changes monotonously to make a complete $2 \pi$ rotation.
We can see from Eq. 4 that varying energy and varying electrostatic potential in some region has similar effects,
specially because winding number is a topological invariant. In other words as energy is varied or $U_1$ is varied,
$\theta_{t_{31}}$, changes monotonously and so the integration in Eq. 61 will not be zero and equality of the
integrands in Eq. 63 cannot be expected. Eq. 62 will still be valid and so while the solid curve in Fig. 8 oscillates
on both sides, the dashed curve oscillates only on one side. Some of the peaks in the solid curve, are very close
to the peaks in the dashed curve but other peaks are reversed in sign. Never the less both the curves are expressing the
same Argand diagram and on careful examination of how the AD behave
in the
$(Im[t_{31}], Re[t_{31}])$ plane with respect to the point (0,0) we can infer the dashed curve is physical meaning LPDOS
is physical. Experimentalists can in fact directly observe the AD \cite{kob} that can help us further to deduce
one curve from the other. That
$\theta_{t_{31}}$, changes monotonously with $U_1$ is shown in Fig. 9.
Since Buttiker et al did not consider Fano resonances, away from the Fano resonances
the negativity of $\rho_{lpd}$ is un-physical and they did not understand Eq. 58 and 59 for the quantum regime. 

\begin{figure}[bt]
\centering
\includegraphics[width=1.1\textwidth, keepaspectratio]{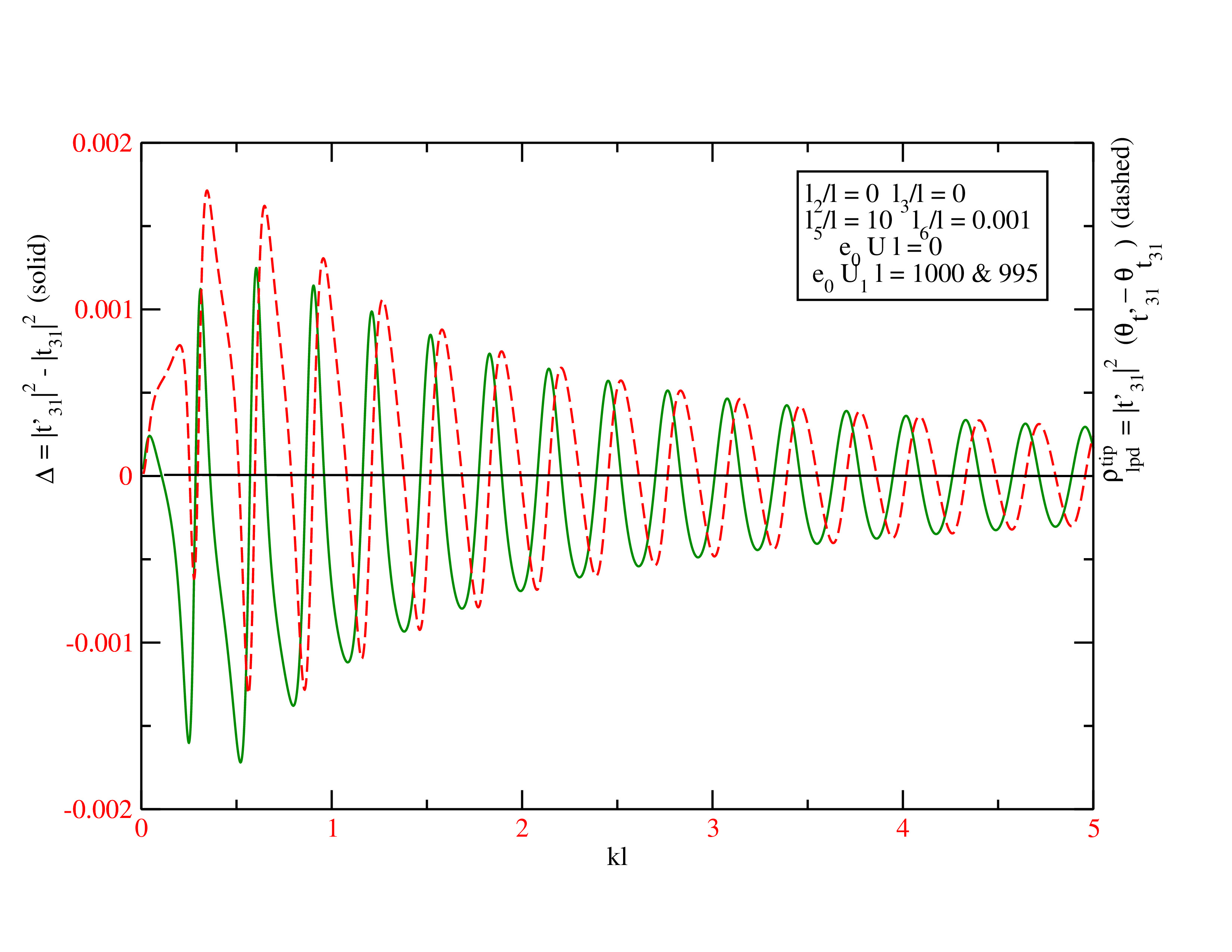}
\captionsetup{labelformat=empty}
\caption{\label{fig33} Fig. 5. In this figure we are plotting the LHS and RHS of Eq. 65 to show that they both can oscillate
identically between positive and negative values. The
primed and un-primed values are for small differences in $U_1$ at a $k$ value that we vary continuously.
The sign change and magnitude of both the curves originate from the smooth cyclic AD of Fig. 4.}
\end{figure}
\begin{figure}[bt]
\centering
\includegraphics[width=1.1\textwidth, keepaspectratio]{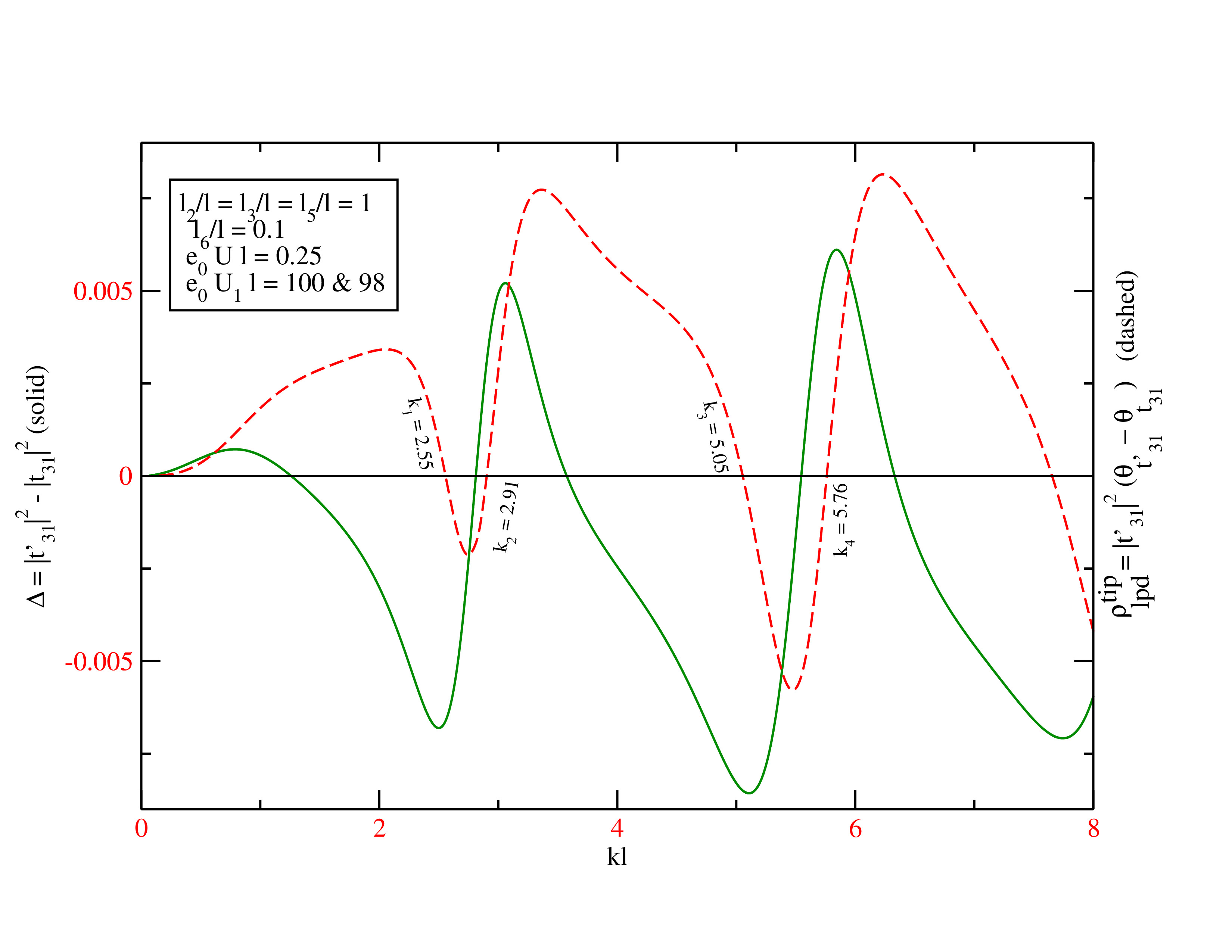}
\captionsetup{labelformat=empty}
\caption{\label{fig33} Fig. 6. In this figure we are plotting the LHS and RHS of Eq. 65
for a different choice of parameters for the system in Fig. 4. The
primed and un-primed values are for small differences in $U_1$ at a $k$ value that we vary continuously.}
\end{figure}
\begin{figure}[bt]
\centering
\includegraphics[width=1.1\textwidth, keepaspectratio]{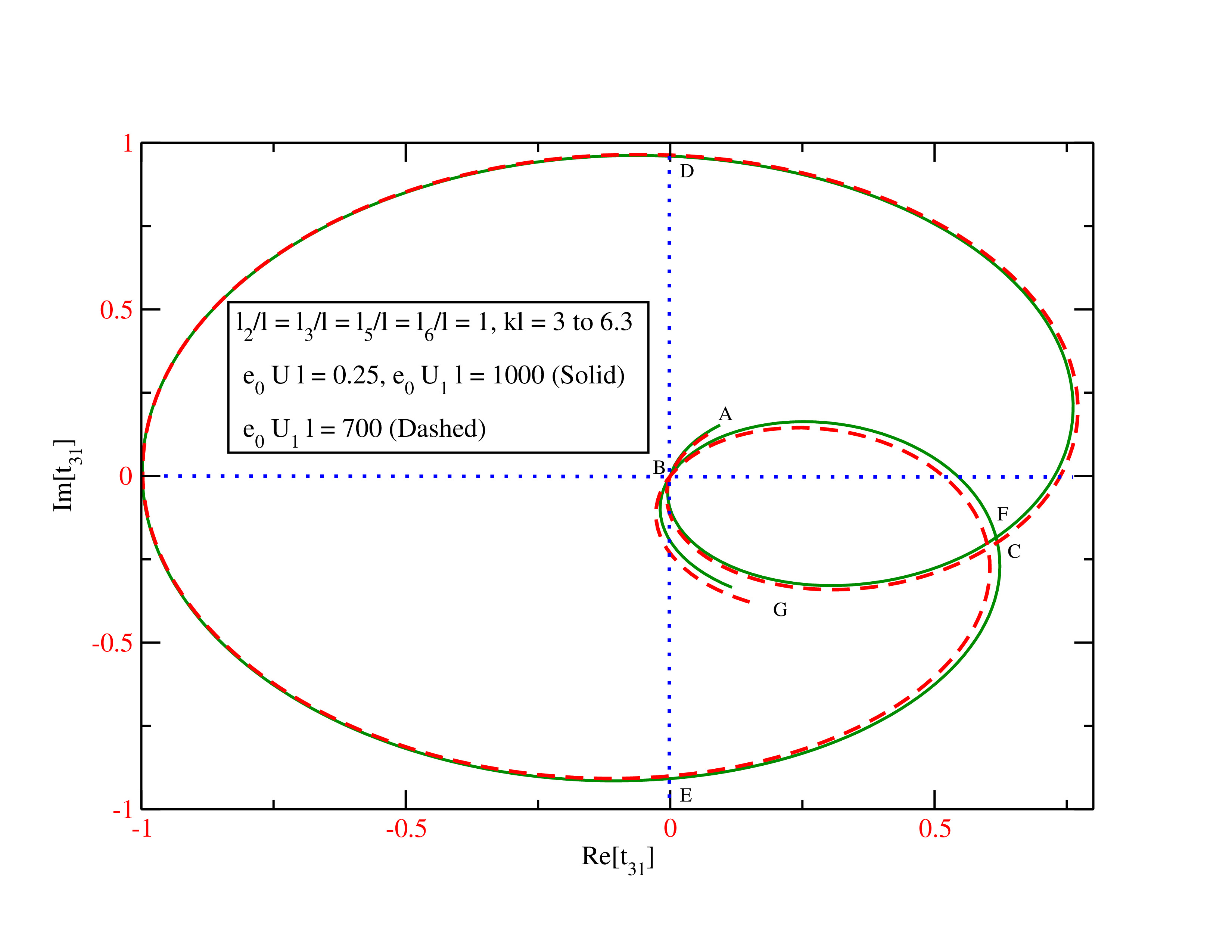}
\captionsetup{labelformat=empty}
\caption{\label{fig33} Fig. 7. In this figure we are plotting the
AD as wave-vector is varied in a certain range for two values of $U_1$ with other parameters fixed and shown in the inset.}
\end{figure}
\begin{figure}[bt]
\centering
\includegraphics[width=1.1\textwidth, keepaspectratio]{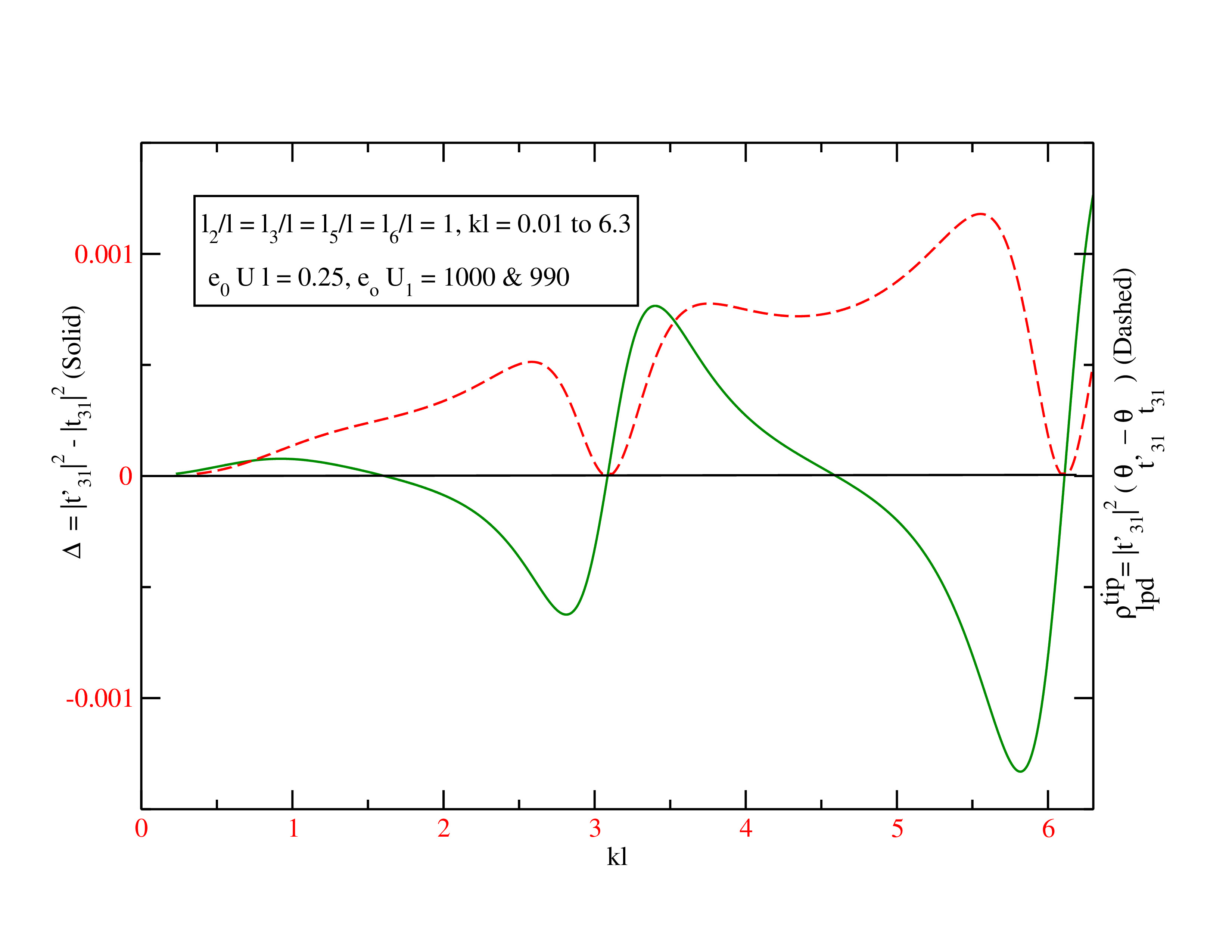}
\captionsetup{labelformat=empty}
\caption{\label{fig33} Fig. 8. In this figure we are plotting the LHS of Eq. 65 (solid curve)
and the RHS of the same equation in dashed curve.
The primed and un-primed values are for small differences in $U_1$ at a $k$ value that we vary continuously.}
\end{figure}
\begin{figure}[bt]
\centering
\includegraphics[width=1.1\textwidth, keepaspectratio]{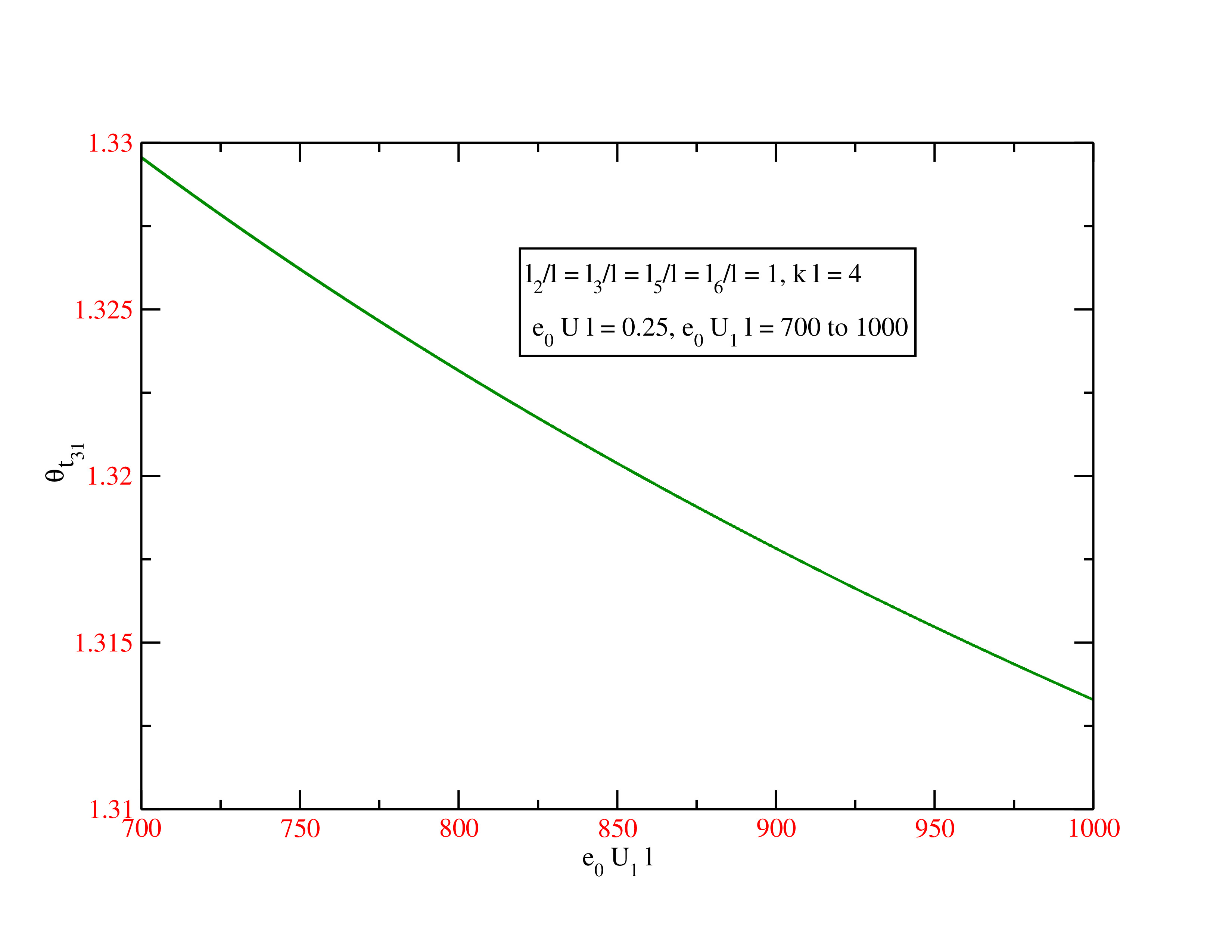}
\captionsetup{labelformat=empty}
\caption{\label{fig33} Fig. 9. In this figure we are
showing that $\theta_{t_{31}}$, changes monotonously with $U_1$ between the solid and dashed curves  in Fig. 7.
Thus, since $\theta_{t_{31}}$ changes monotonously with $U_1$ as well as with
wave-vector, then
$(\theta_{t'_{31}} - \theta_{t_{31}})$ can only vary on one side of the horizontal axis.
Real part of $t_{31}$
or imaginary part of $t_{31}$ can change sign as wave-vector is varied but
$\theta_{t_{31}}$ changes monotonously with $U_1$ in the wave-vector range considered in Fig. 8.}
\end{figure}

\section{Conclusion}

AD of a scattering matrix element like
$t_{31}$ or $s_{\alpha \gamma}$ can show two kinds of topological behavior. In one case the AD closes on one side of the
complex plane
$(Im[t_{31}], Re[t_{31}])$ without enclosing the singularity (0,0). In that case both
$\rho_{lpd}$ and
$|t'_{31}|^2 - |t_{31}|^2$ will oscillate with positive-negative values and will be equal in magnitude.
Thus the equality derived in Ref. \cite{gra} make sense and
$|t'_{31}|^2 - |t_{31}|^2$ is determined by
$\rho_{lpd}$. However, if the AD of
$t_{31}$ does enclose the singularity in the plane
$(Im[t_{31}], Re[t_{31}])$ then the equality of the integrations in Eq. 63 itself breaks down because of topological reasons.
Then the integrands will still be comparable in magnitudes as derived in Ref. \cite{gra} but they will be qualitatively different,
not resembling each other. The LHS of Eq. 63 not being a winding number will always be zero while the RHS being a winding number will
be unity. Never the less, 
$\rho_{lpd}$ still determines
$|t'_{31}|^2 - |t_{31}|^2$ as they are both expressions of the same AD, however
$\rho_{lpd}$ will oscillate on the positive side only while
$|t'_{31}|^2 - |t_{31}|^2$ will oscillate with positive and negative values. One can experimentally determine this AD \cite{kob} and
by carefully examining the AD one can infer the behavior of
$\rho_{lpd}$ and
$|t'_{31}|^2 - |t_{31}|^2$ and relate them. However, it is easy to tune a mesoscopic system into a regime with only Fano resonances
and thereby obtain an equality between
$\rho_{lpd}$ and
$|t'_{31}|^2 - |t_{31}|^2$ for any system.
Thus in principle a measurable quantity
$|t'_{31}|^2 - |t_{31}|^2$ is determined by a number of local partial states
$\rho_{lpd}(\textbf r)$ in spite of all possible linear superposition of states at the point $\textbf r$ is demonstrated
and hence solves the quantum measurement problem in the extreme quantum regime as well.
Since $\rho_{lpd}(\textbf r)$ is thereby real, and is related to a local time that is fully consistent with relativity as
well as quantum mechanics, hence unifies the two. Prof. Roger Penrose has been repeatedly stating in his public talks that
since outcome of a quantum measurement within the axiomatic framework of quantum mechanics is random it will not be possible to
unify relativity and quantum mechanics unless one can show that quantum measurements are deterministic.
That is basically because if the theory of quantum mechanics fail to give a mechanism for measurement outcomes at a regime where it
works the best, that is the micro world, then it cannot give a measurement outcome in the limit when the macro world
evolves from the micro world.
Our solution to the problem of quantum measurement and unification of classical and quantum worlds corroborates Penrose's intuition.
The $\rho_{lpd}$ cannot be defined in the axiomatic framework of quantum mechanics, but can deterministically give
the outcome of a quantum measurement, and naturally solve the problem of unification.

\clearpage
 
\bibliographystyle{References}

\end{document}